\documentclass[runningheads]{llncs}
\usepackage[T1]{fontenc}
\usepackage{graphicx}
\usepackage{caption}
\usepackage{subcaption}
\usepackage{multirow}
\usepackage{siunitx}
\usepackage[export]{adjustbox}
\usepackage{subcaption}
\usepackage[utf8]{inputenc} 
\usepackage[T1]{fontenc}    
\usepackage{hyperref}       
\usepackage{url}            
\usepackage{booktabs}       
\usepackage{amsfonts}       
\usepackage{nicefrac}       
\usepackage{microtype} 
\usepackage{xcolor,colortbl}         
\usepackage{graphicx} 
\usepackage{varwidth}
\usepackage{wrapfig}
\usepackage{acronym}
\usepackage{xspace}
\usepackage{tikz}
\usepackage{multirow}
\usepackage[export]{adjustbox}
\usepackage{graphicx}
\usepackage{amsmath}
\usepackage{algorithm}
\usepackage{algpseudocode}
\usepackage{hyperref}

\newcommand{\ud}{\mathrm{d}}
\newcommand{\bfx}{\mathbf{x}}
\newcommand{\bff}{\mathbf{f}}
\newcommand{\bfw}{\mathbf{w}}

\begin{document}

\acrodef{WSI}{Whole Slide Image}
\acrodef{CDM}{Cascaded Diffusion Model}
\acrodef{URCDM}{Ultra-Resolution Cascaded Diffusion Models}
\acrodef{LRDM}{Lower Resolution Diffusion Model}
\acrodef{MAE}{Mean Absolute Error}
\acrodef{GAN}{Generative Adversarial Networks}
\acrodef{FID}{Fréchet Inception Distance}
\acrodef{IP}{Improved Precision}
\acrodef{IR}{Improved Recall}

\title{
URCDM: Ultra-Resolution Image Synthesis in Histopathology}
%
%
\author{
Sarah Cechnicka \inst{1}
\and
James Ball \inst{1}
\and
Matthew Baugh \inst{1}
\and
Hadrien Reynaud \inst{1}
\and
Naomi Simmonds \inst{2} 
\and
Andrew P.T. Smith \inst{2} 
\and
Catherine Horsfield \inst{2} 
\and
Candice Roufosse \inst{2} 
\and
Bernhard Kainz \inst{1,3}
}

\authorrunning{S. Cechnicka et al.}
%
\institute{Departament of Computing, Imperial College London, UK 
\and
NHS Trust, London, UK
\and
Friedrich–Alexander University Erlangen–N\"urnberg, DE
\\
\email{sc7718@imperial.ac.uk}\\
}

\maketitle              

\begin{abstract}
Diagnosing medical conditions from histopathology data requires a thorough analysis across the various resolutions of Whole Slide Images (WSI). However, existing generative methods fail to consistently represent the hierarchical structure of WSIs due to a focus on high-fidelity patches. To tackle this, we propose Ultra-Resolution Cascaded Diffusion Models (URCDMs) which are capable of synthesising entire histopathology images at high resolutions whilst authentically capturing the details of both the underlying anatomy and pathology at all magnification levels. We evaluate our method on three separate datasets, consisting of brain, breast and kidney tissue, and surpass existing state-of-the-art multi-resolution models.
Furthermore, an expert evaluation study was conducted, demonstrating that URCDMs consistently generate outputs across various resolutions that trained evaluators cannot distinguish from real images. All code and additional examples can be found on \href{https://github.com/scechnicka/URCDM}{GitHub}. 
\end{abstract}

\section{Introduction}

\noindent\hspace{-0.6em}
Professional pathologists rely on both low and high-magnification views to make critical decisions. At lower magnifications, they assess overall tissue architecture to identify areas of interest, such as tumour distribution.
Medium magnifications allow for closer examination of smaller tissue structures and cell groupings, providing details on cellular changes.
At the highest magnifications, individual cell morphology is scrutinized for specific abnormalities crucial for accurate diagnosis. Each level of magnification contributes essential information, enabling a comprehensive pathological assessment.

Machine learning techniques~\cite{Wu2023Application,deephistopatho} could partially automate such diagnosis, but, unfortunately, access to \ac{WSI} data from real patient populations is often restricted due to privacy concerns, making it difficult to develop robust models. 
Diffusion models have demonstrated significant value in this area, facilitating the creation of synthetic data for medical education and the development of machine learning models for automated analysis, all while ensuring the protection of patient privacy.

In histopathology, where staining differences hinder generalisability, diffusion models achieve a more dependable performance across diverse data sources~\cite{Cechnicka2023}. 
Most research focuses on the use of generative approaches at small-scale patch levels, where these techniques excel at generating local, high-fidelity details~\cite{ganreview,macenko,Federated}.
However, significant challenges arise when transitioning from patch-level analysis to \ac{WSI}, such as memory constraints, long sampling times, and a lack of training data. Additionally, these methods risk data curation biases, as training and validation datasets for patch-level problems are often compiled by the same experts or under identical guidelines, potentially skewing task outcomes~\cite{Ciga123}.
Many existing methods are not applicable for large context images, as they were originally designed for much lower resolutions~\cite{NASDM,morphdiff}. This transition introduces complexities related to preserving contextual information and ensuring seamless integration between different magnification levels.
The development of tailored algorithms capable of handling such scale is therefore crucial.
This is particularly important as accurate generation and analysis at the \ac{WSI} level increases the number and variety of available data  samples.
This also enables the use of more complex downstream algorithms, which operate on the entire image at different scales, \emph{e.g.}, You Only Look Twice~\cite{yolt}.

\noindent\textbf{Contributions.}
We introduce a method leveraging \ac{URCDM}s to produce high-fidelity, photo-realistic histopathology images at the \ac{WSI} scale, marking a first in the field. Our approach uniquely captures detailed features at multiple magnifications and supports long-range contextual understanding, overcoming the memory limitations of attention -based models. It achieves this at a significantly reduced computational cost, facilitating efficient image generation even in data-intensive \ac{WSI} learning contexts. 

\noindent\textbf{Related Work.}
The strategy of enhancing datasets with artificially created images has been a long-standing idea, yet its realisation evolved slowly~\cite{databalance,InsMix,gupta2019gan}.
Techniques based on auto-encoders~\cite{kingma2013auto} or \ac{GAN}s~\cite{goodfellow2020generative} often exhibited significant limitations, mainly due to a noticeable domain gap between real and synthetic images.
This makes machine learning on enriched data difficult since it is often easier for downstream application models, like pathology detection, to discriminate data sources over desired training signals, which introduces significant confounders. 
This is further exasperated when dealing with gigapixel images like \ac{WSI}, when more information is handled and a bottom-up or top-down approach through patching is required.

Diffusion models have been shown to produce synthetic images that better match the distribution of target datasets~\cite{ramesh_zero-shot_2021,rombach_high-resolution_2022,imagen}.
Recent applications of \ac{CDM}s~\cite{imagen} have improved multi-scale information integration in medical imaging, pushing the boundaries of diagnostic accuracy and patient care~\cite{reynaud2023featureconditioned}.
These diffusion models are chained together, first generating a low-resolution image, which gets repeatedly upsampled by each model in the cascade.
Compared to attempting to train one model directly at the target resolution, this process allows for smaller models and parallel training, making these large-scale experiments much more tractable in time,compute resources and allows for seamless edges.

\section{Method}

Diffusion models have emerged as powerful generative models, capable of being parameterized through various approaches to accurately model complex data distributions.
In this work, we follow the Stochastic Differential Equation definition \cite{scorebasedgen}. 
We define our forward (Eq. \ref{eq:forward_sde}) and backward (Eq. \ref{eq:backward_sde}) processes as:
\begin{align}
    \ud \bfx _{+} &= \bff(\bfx, t) \ud t + {g}(t) \ud \bfw, \label{eq:forward_sde}\\
    \ud \bfx _{-} &= [\bff(\bfx, t) - g(t)^2  \nabla_{\bfx}  \log p_t(\bfx)] \ud t + g(t) \ud \bar{\bfw},\label{eq:backward_sde}
\end{align}
where $\ud \bfx _{+}$ is the noise to be added to the sample, while $\ud \bfx _{-}$ is the noise to be removed from the sample, following the epsilon (noise prediction) objective. 
$\bfw$ and $\bar{\bfw}$ are forward and backward Wiener processes, respectively introducing stochasticity. 
$\bff(\cdot, t)$ is the drift coefficient that models the deterministic part of the state's change in the diffusion process over time. 
$g(\cdot)$ is the diffusion coefficient of $\bfx(t)$. 
$\nabla_{\bfx}  \log p_t(\bfx)$ is the score function we learn with a neural network. 
We use this parametrisation for all our diffusion models.

\ac{CDM}s work by first generating a low-resolution image $I_0$ with a base model $C_{\phi_0}$ from Gaussian noise.
$I_0$ is subsequently used as a conditional input for a second diffusion model $C_{\phi_1}$\cite{ondistillationofguideddiffusionmodels}, which generates an image of higher resolution $I_1$, that is based on $I_0$.
Generally, the nth super-resolution stage will condition on the lower-resolution image $I_{n-1}$ generated in the previous stage.
We can write 
\begin{equation}
C_{\phi_n}\left(\sigma, I_{n-1}\right)=F_{\phi_n}\left(c_{\text {noise }}(\sigma), I_{n-1}\right), C_{\phi_0}\left(\sigma\right)=F_{\phi_0}\left(c_{\text {noise }}(\sigma)\right).
\end{equation}

\begin{figure}[ht]
\centering
\includegraphics[width=\linewidth]{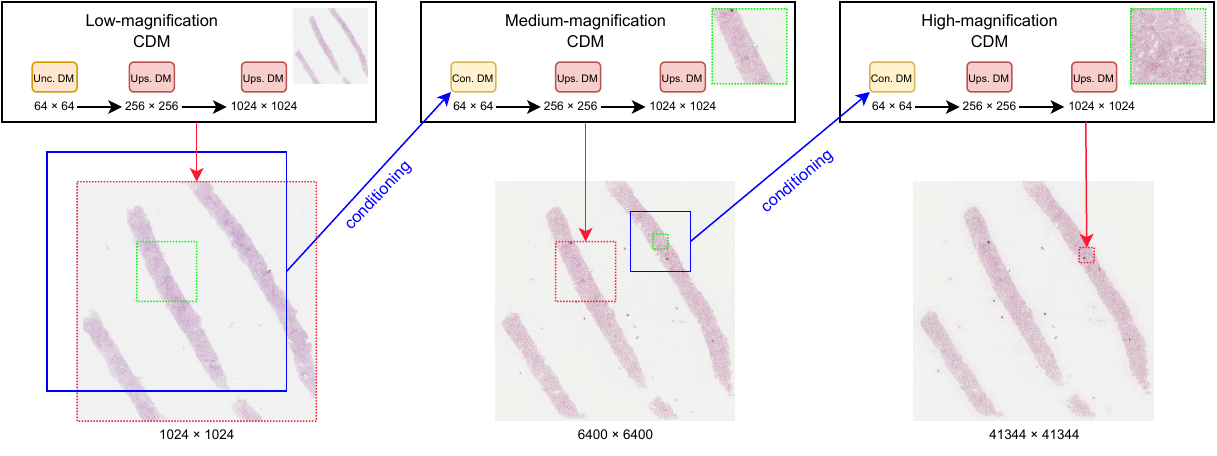}
\caption{Detailed overview of the \ac{URCDM} image generation process. The medium and high-magnification \ac{CDM}s are sampled many times, and the patches generated are stitched together; one sample is shown as an example. A \textcolor{blue}{blue} outline indicates \textcolor{blue}{ the lower-magnification conditioning image}, to teach the context for the new generation process. A \textcolor{green}{green} outline indicates \textcolor{green}{the resultant patch} that will be `zoomed in' on and generated by a baseline \ac{CDM}. \textcolor{red}{Red} lines indicate \textcolor{red}{the output of each magnified image}. Not to scale.}
\label{fig:urcdm-gen-process}
\end{figure}

\ac{URCDM}s extend this principle further.
The overall pipeline consists of three stages which form a cascade, with each outputting a \ac{WSI} at higher and higher resolutions, as shown in Figure~\ref{fig:urcdm-gen-process}.
Within each stage there is a \ac{CDM}, which aims to upsample a specific region of interest.
Every \ac{CDM} starts by generating an image of dimension $64\times 64$ and then upscales it twice, first to $256\times 256$ and then to $1024 \times 1024$.

The first stage directly uses a standard \ac{CDM} to output an entire \ac{WSI} of size $\num{1024} \times \num{1024}$.
The second stage then uses a separate \ac{CDM} to upsample patches of this image using a combination of conditioning and inpainting.
We adapt the \ac{CDM} so that each stage is also conditioned on a $\num{1024} \times \num{1024}$ patch of the image from the previous stage, centered on the patch being upsampled.
This allows it to see both the region for upsampling and its surrounding context, ensuring that the generated patch is consistent with the high-level structure of the image.
Our second adaptation starts with how we use the \ac{CDM} to generate a grid of overlapping patches.
When generating a new patch, using inpainting, any pixels in regions that have already been generated as part of neighbouring patches are kept constant throughout the reverse diffusion process.
Giving the diffusion model this information enables it to produce patches that seamlessly blend together without any artifacts between patch boundaries.
At the end of stage 2 all the upsampled patches are combined together to make a \ac{WSI} of size $\num{6400} \times \num{6400}$.

Stage 3 then follows the same structure as stage 2, except now the images used for conditioning are taken from the output of stage 2, resulting in the final \ac{WSI} having a resolution of $\num{41344} \times \num{41344}$.

In our proposed \ac{URCDM}, there are nine diffusion models in total that form three separate \ac{CDM}s. All models are trained independently and in parallel on nine separate Nvidia A100 GPUs, which takes 24 hours for 200,000 steps. 
During inference a low-quality \ac{WSI} is generated with the use of the low-resolution \ac{CDM} with size $\num{1024} \times \num{1024}$. Overlapping patches of this generated low-resolution image are then used to condition the second \ac{CDM}, by giving it the spatial context needed for generation. The second model generates images of size $\num{1024} \times \num{1024}$ for the centre or each conditioning patch. After stitching them together, and accounting for overlaps, the resolution of the \ac{WSI} increases to $\num{6400} \times \num{6400}$. This process is repeated for the final high-resolution model, yielding a final synthetic \ac{WSI} of $\num{41344} \times \num{41344}$ pixels, which takes approximately 24h on nine separate Nvidia A100 GPUs to generate. The size of the medium-magnification \ac{CDM} was chosen to be approximately halfway between low-resolution and high-resolution magnifications. The \ac{URCDM} is not restricted to these resolutions, which can trivially be changed to suit other datasets.

\noindent\textbf{Implementation details:} We base on Imagen~\cite{imagen} using the \verb|imagen-pytorch|~\cite{imagenpytorch} library. Each \ac{CDM} targets a different magnification of the overall image. 
Gradient clipping was implemented (set to 1) as well as v-parametrisation, to avoid lower quality fine details, blurring or heavy distortion of lower magnification images when `zooming in'. Inpainting~\cite{cheng2022inout} is used to smoothly merge generated patches together with minimal seams. The resulting dependency between patches is especially important when batch-processing images or sampling on multiple GPUs, as it determines when patches can be generated in parallel. This ensures reasonable sampling speeds. 12.5\% of overlap (both in the vertical and horizontal directions) was chosen for our experiments to minimize patching artefacts while reasonably decreasing the total number of patches generated. High-magnification patches of the whole slide image that are mostly white are ignored in both training and sampling. Instead, all white patches are replaced with an upscaled version of the medium-magnification image. All images were cropped or padded to $\num{41300} \times \num{41300}$ pixels. 

\section{Evaluation and Results} 

\noindent\textbf{Datasets and Preprocessing:} We explore our approach on three datasets for evaluation. The first dataset is the public low-grade glioma dataset from The Cancer Genome Atlas (TCGA) archive \cite{datasettcga} containing 344 \ac{WSI}s, hereafter referred to as GLIOMA. The second is a public The Cancer Genome Atlas Breast Invasive Carcinoma (TCGA-BRCA) dataset \cite{datatcga_BRCA}. The dataset comprises \num{1978} high-resolution \ac{WSI}s of various tissues that have been stained using a variety of protocols, designated as BREAST. The final dataset is a proprietary collection of 428 Kidney Transplant Pathology WSI slides with an average resolution of $\num{40000} \times \num{40000}$ per slide, named KIDNEY.  
Ultra-resolution images must consistently appear realistic at multiple scales both when ‘zoomed-out’ and when ‘zoomed-in'. Lower-resolution patches generated by a \ac{LRDM} are evaluated against StyleGAN3~\cite{stylegan3}, and the Morphology Focused Diffusion Probabilistic Model (MorphDiff) (~\cite{morphdiff}. To accommodate the size limitations of the models, which handle images up to $1024\times1024$, all \ac{WSI}s were segmented into patches. These patches were then resized to $64\times64$ for the initial diffusion model training, augmented with basic transformations, and further scaled to $256\times256$ for the first super-resolution model. An \ac{LRDM} 
has been chosen as the backbone for outpainting for fair comparison~\ref{tab:fid-results}. 

Secondly, long-distance spatial coherency is evaluated. The realism of high-resolution images is compared against baseline unconditional diffusion models generating high-resolution images using outpainting.  All datasets used for experimenting on \ac{WSI} level are first cropped to patches size $\num{40000}\times\num{40000}$ pixels. These images are then downsampled to $\num{1024}\times\num{1024}$ for the low magnification models and undergo the same transformations and resizing to $256\times256$ and $64\times64$ as in the low-resolution models for training. The $\num{40000}\times\num{40000}$ patches are then separately resized to $\num{6400}\times\num{6400}$ and cropped to $\num{1024}\times\num{1024}$ for the mid-magnification model 
and undergo the same training preprocessing as the \ac{LRDM}. Finally, for the high-magnification model the  $\num{40000}\times\num{40000}$ patches are cropped to $\num{1024}\times\num{1024}$ patches that then follow the same training preprocessing as all other models. 

\noindent\textbf{FID and pFID:} The \ac{FID}~\cite{fid} is a metric to evaluate the image quality and diversity of generated images by comparing two Gaussian distributions $\mathcal{N}(\boldsymbol{\mu}_r, \boldsymbol{\Sigma}_r)$, and $\mathcal{N}(\boldsymbol{\mu}_g, \boldsymbol{\Sigma}_g)$, where $r$ and $g$ represent real and generated images respectively. $\boldsymbol{\mu}$ and $\boldsymbol{\Sigma}$ are found by fitting to latent-space feature vectors of real images and generated images~\cite{fid}.
FID becomes impractical for very high-resolution images due to computational constraints and the need to capture local image quality across vast areas. 
Patch-FID (pFID), introduced by~\cite{anyresgan} is a variation of \ac{FID}~\cite{fid} for very high-resolution images and is used for the evaluation of gigapixel performance. It works by taking random crops at random scales from the image and computing the FID between the real and generated patches by sampling different patches at matching scales and positions~\cite{anyresgan}.

\noindent\textbf{IP and IR:} \ac{IP}~\cite{Kynkaanniemi2019} quantifies the proportion of generated samples that align with the real data manifold.
Conversely, \ac{IR} measures the extent to which the real data manifold is encapsulated within the synthetic data manifold, reflecting the model's ability to capture data diversity. These metrics are derived by computing the k-Nearest Neighbors distances for each sample, facilitating a non-parametric approximation of the data manifolds. While FID is a robust and popular metric for comparing the overall similarity of real and generated data distributions, \ac{IR} and \ac{IP} offer a more detailed, balanced assessment of a generative model's ability to produce diverse, high-quality samples that closely match the characteristics of the real data.

\noindent\textbf{Metric Implementation:} To accurately model \ac{URCDM}'s capabilities various post-processing steps were meticulously followed. For pFID-50k we utilized various crops at different magnifications, all resized to $\num{1024}\times\num{1024}$ pixels. These were kept consistent with the training dataset for each cascading level, facilitating an accurate measure of the generated images' fidelity compared to the real images. 
Given the computational intensity of our method, especially at higher magnifications, we use IP and IR metrics on approximately \num{10000} patches for each magnification level, except for magnification 0, which was limited to the real samples available. For magnifications 1 and 2, the analysis was based on roughly \num{10000} patches randomly sampled from about \num{800} generated images. Although this limited the performance on both magnifications, this decision was necessary as generating a single full-size \ac{WSI} at magnification~2 requires approximately 24 hours on 8 NVIDIA A100 GPUs. The values in Table \ref{tab:fid-results} are then uniformly averaged over these three performance measurements.

\begin{table}[t]
\caption{Results on the unconditional generative model and baselines trained on all datasets report FID-10k, IP and IR on 10k image of $\num{1024}\times\num{1024}$ patches (MorphDiff~\cite{morphdiff} uses $128\times128$). Results of pFID-50k, IP and IR on different methods of generating ultra-resolution images using diffusion models on $\num{1024}\times\num{1024}$ patches.}
\label{tab:fid-results}
\resizebox{\linewidth}{!}{
\begin{tabular}{c p{7em} p{4em} p{5em} p{2em} p{2em} p{4em} p{5em} p{2em} p{2em} p{4em} p{5em} p{2em} p{2em}}

\toprule
    & & \multicolumn{4}{c}{\textbf{\textbf{KIDNEY}}} & \multicolumn{4}{c}{\textbf{\textbf{GLICOMA}}} &\multicolumn{4}{c}{\textbf{\textbf{BREAST}}} \\

    \cmidrule(lr){3-6}\cmidrule(lr){7-10}\cmidrule(lr){11-14}
    &\multirow{2}{*}{Model} &
      \multicolumn{2}{c}{Fidelity $\downarrow$} &
      \multirow{2}{*}{IP $\uparrow$} &
      \multirow{2}{*}{IR $\uparrow$} &
      \multicolumn{2}{c}{Fidelity $\downarrow$} &
      \multirow{2}{*}{IP $\uparrow$} &
      \multirow{2}{*}{IR $\uparrow$}  &
      \multicolumn{2}{c}{Fidelity $\downarrow$} &
      \multirow{2}{*}{IP $\uparrow$} &
      \multirow{2}{*}{IR $\uparrow$}  \\
      && \multicolumn{1}{c} {$FID_{10k}$} & {$pFID_{50k}$} &  & 
      &  {$FID_{10k}$} & {$pFID_{50k}$} &  & 
      &  {$FID_{10k}$} & {$pFID_{50k}$} & \\
      \cmidrule(lr){2-2} \cmidrule(lr){3-6}\cmidrule(lr){7-10}\cmidrule(lr){11-14}
      
\multirow{3}{*}{\rotatebox[origin=c]{90}{\tiny{patch}}}&StyleGAN3~\cite{stylegan3} & \multicolumn{1}{c}{38.62}& \multicolumn{1}{c}{n.a} & \multicolumn{1}{c}{0.74}&\multicolumn{1}{c}{0.08} & \multicolumn{1}{c}{\textbf{48.59}} &\multicolumn{1}{c}{n.a}&\multicolumn{1}{c}{\textbf{0.72}}&\multicolumn{1}{c}{0.08}& \multicolumn{1}{c}{-} &\multicolumn{1}{c}{n.a}&\multicolumn{1}{c}{-}&\multicolumn{1}{c}{-}\\
&Morph-Diff~\cite{morphdiff}&\multicolumn{4}{c}{\cellcolor{white}NO PUBLIC CODE}& \multicolumn{1}{c}{20.11} &\multicolumn{1}{c}{n.a}&\multicolumn{1}{c}{0.26}&\multicolumn{1}{c}{0.85} & \multicolumn{4}{c}{\cellcolor{white}NO PUBLIC CODE}\\

&\ac{LRDM} &\multicolumn{1}{c} {\textbf{10.35}} & \multicolumn{1}{c}{n.a} & \multicolumn{1}{c}{\textbf{0.82}}&\multicolumn{1}{c}{\textbf{0.57}}&\multicolumn{1}{c}{34.31}&\multicolumn{1}{c}{n.a}&\multicolumn{1}{c}{0.70}&\multicolumn{1}{c}{\textbf{0.85}} & \multicolumn{1}{c}{66.17} &\multicolumn{1}{c}{n.a}&\multicolumn{1}{c}{0.91}&\multicolumn{1}{c}{0.30}\\

\cmidrule(lr){2-2}\cmidrule(lr){3-6}\cmidrule(lr){7-10}\cmidrule(lr){11-14}
    
\multirow{2}{*}{\rotatebox[origin=c]{90}{\tiny{WSI}}}&Outpainting~\cite{cheng2022inout} &\multicolumn{1}{c}{n.a}&\multicolumn{1}{c}{150.15}&\multicolumn{1}{c}{0.63}&\multicolumn{1}{c}{0.13}& \multicolumn{1}{c}{n.a} &\multicolumn{1}{c}{192.89}&\multicolumn{1}{c}{0.60}&\multicolumn{1}{c}{0.23} &\multicolumn{1}{c}{n.a}&\multicolumn{1}{c}{166.24} &\multicolumn{1}{c}{0.56}&\multicolumn{1}{c}{0.20}\\  
&\ac{URCDM} (ours) &\multicolumn{1}{c}{n.a}&\multicolumn{1}{c}{\textbf{39.52}} &\multicolumn{1}{c}{\textbf{0.70}}&\multicolumn{1}{c} {\textbf{0.18}}& \multicolumn{1}{c}{n.a} &\multicolumn{1}{c}{\textbf{52.38}}&\multicolumn{1}{c}{\textbf{0.76}}&\multicolumn{1}{c}{\textbf{0.51}}& \multicolumn{1}{c}{n.a}&\multicolumn{1}{c}{\textbf{67.66}} &\multicolumn{1}{c}{\textbf{0.69}}&\multicolumn{1}{c}{\textbf{0.31}}\\ 
\bottomrule
\end{tabular}
}
\end{table}

\begin{table}[htbp]
\centering
\begin{minipage}[t]{0.58\linewidth} 
\caption{Human expert evaluation. 
TP = true positives, FP = false positives. $p$ = proportion of incorrectly classified samples. The final averaged total values in bold for the expectation corrected proportion $|p-0.5|$.}
\label{tab:human-eval}%
\resizebox{\linewidth}{!}{%
\begin{tabular}{ccccccccccc}
\toprule
\bfseries User & &TP & FP & $p$ & \bfseries {$|p-0.5|$ } & & TP & FP & $p$ & \bfseries {$|p-0.5|$ }  \\ \midrule
Pathol. 1 &  \multirow{5}{*}{\rotatebox{90}{\ac{LRDM}}}
& 250  & 179    & 0.4172    & 0.0823 & 
\multirow{5}{*}{\rotatebox{90}{\ac{URCDM}}} 
&  61   & 66    & 0.5197    & 0.0197  \\  
Pathol. 2 && 106    & 145   & 0.5777  & 0.0777 &  
& 152  & 6   & 0.0380  & 0.4620 \\ 
Pathol. 3 && 29     & 99   & 0.7734   & 0.2734 &  
& 28   & 33   & 0.5410   & 0.0410  \\ 
Pathol. 4 & &-&-&-&- & 
&  47   & 10   & 0.1754   & 0.3246  \\ 
Non-expert    && 110  & 162    & 0.5956      & 0.0956 & 
& 29  & 21    & 0.4200   & 0.0800\\ \midrule
Total && 495 & 585 & 0.5417 & \bfseries {0.1074}& 
&317 & 136 & 0.3002 & \bfseries {0.2219}  \\ 
\bottomrule
\end{tabular}
}
\end{minipage}
\hfill
\begin{minipage}[t]{0.38\linewidth} 
\caption{Detailed average IP and IR metrics for $1024\times1024$ patches across three magnification levels for all three datasets. 
}
\label{tab:ipir}
\resizebox{\linewidth}{!}{%
\begin{tabular}{lcccccc}

\toprule
 \multirow{2}{*}{Model} &
      \multicolumn{2}{c}{mag. 0} &
      \multicolumn{2}{c}{mag. 1} & 
      \multicolumn{2}{c}{mag. 2} \\
      & IP$\uparrow$ & {IR}$\uparrow$ & {IP}$\uparrow$ & {IR}$\uparrow$
      &  {IP}$\uparrow$ & {IR} $\uparrow$  \\ \midrule
KIDNEY  & 0.87&0.42&0.60&0.10& 0.62 & 0.03 \\ 
BREAST  & 0.54&0.45&0.70&0.20&0.83 & 0.27\\ 
GLICOMA & 0.76&0.63&0.61&0.40&0.91& 0.50\\ 
\bottomrule
\end{tabular}
}
\end{minipage}
\end{table}

\noindent\textbf{Results:} 
The quantitative results of the experiments can be seen in Tables~\ref{tab:fid-results}, \ref{tab:ipir}. Table \ref{tab:fid-results} provides a general comparison with other state-of-the-art models whereas Table~\ref{tab:ipir} shows detailed generative results for our model across 3 magnification scales.  Figure~\ref{fig:urcdm-kidney} shows randomly selected example images generated by the \ac{URCDM} and comparison methods. \ac{URCDM}s consistently outperform all state-of-the-art methods on all three datasets. 
Generally, \ac{URCDM}s exhibit considerable plausible diversity across different regions at high magnification levels while preserving a high quality. At lower magnification levels, the images accurately represent the fundamental structures of the \ac{WSI}.

\newcommand\x{2.9cm}
\begin{figure}[tb]
\centering
\begin{tabular}[c]{p{0.1cm}ccp{0.1cm}cc}
\centering 
   & \tiny{Outpainting} & \tiny{\ac{URCDM}} && \tiny{StyleGAN} & \tiny{\ac{URCDM}}   \\ 
    \raisebox{1.0cm}{\rotatebox[origin=l]{90}{\tiny{KIDNEY}}}
    & \includegraphics[height=\x]{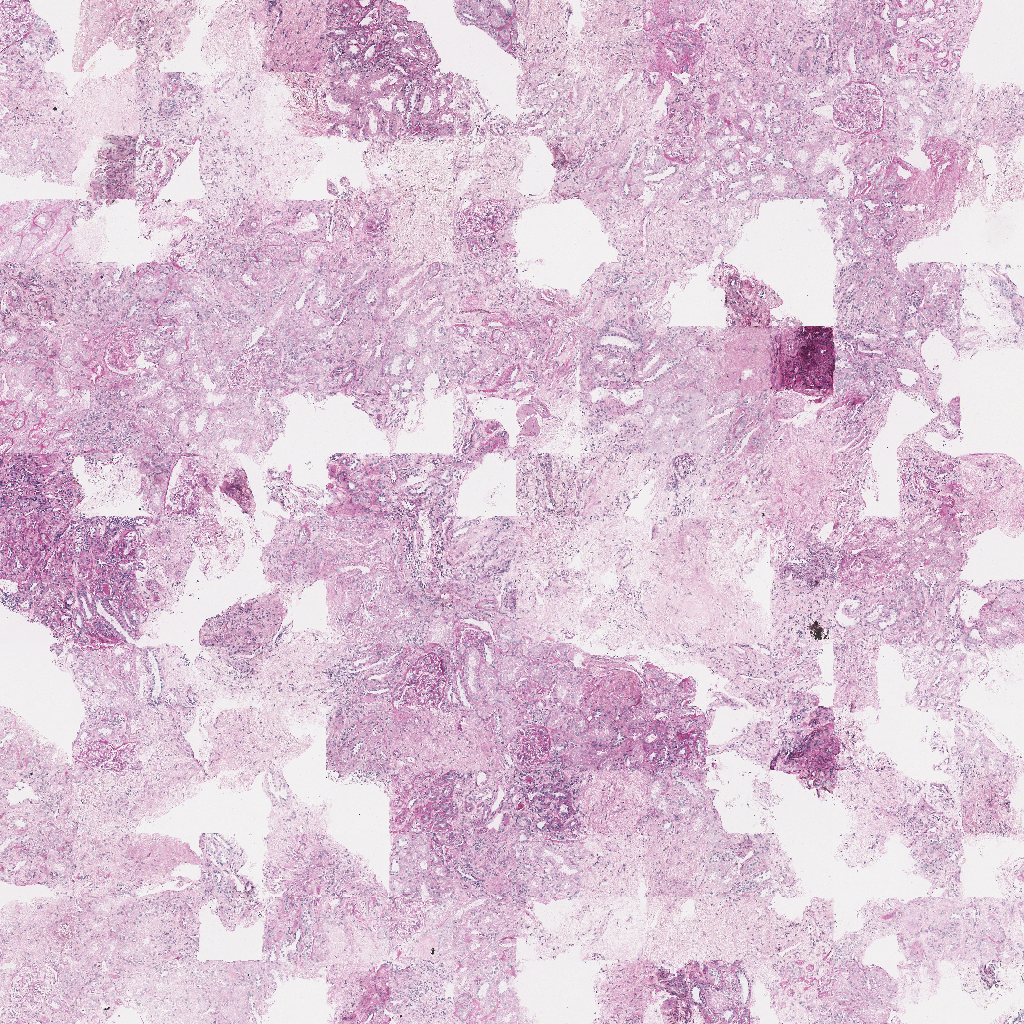}&
    \includegraphics[height=\x]{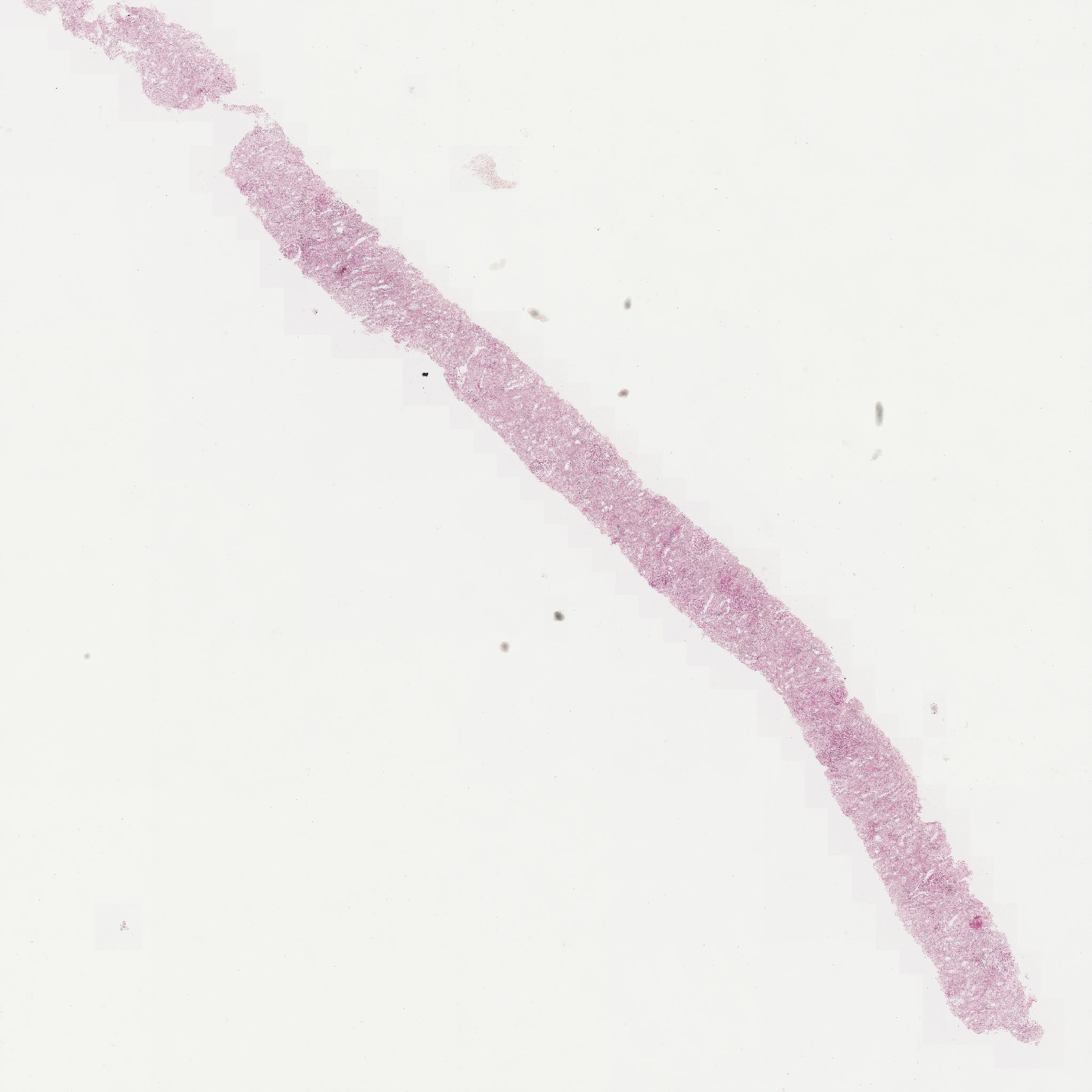}& &
    \includegraphics[height=\x]{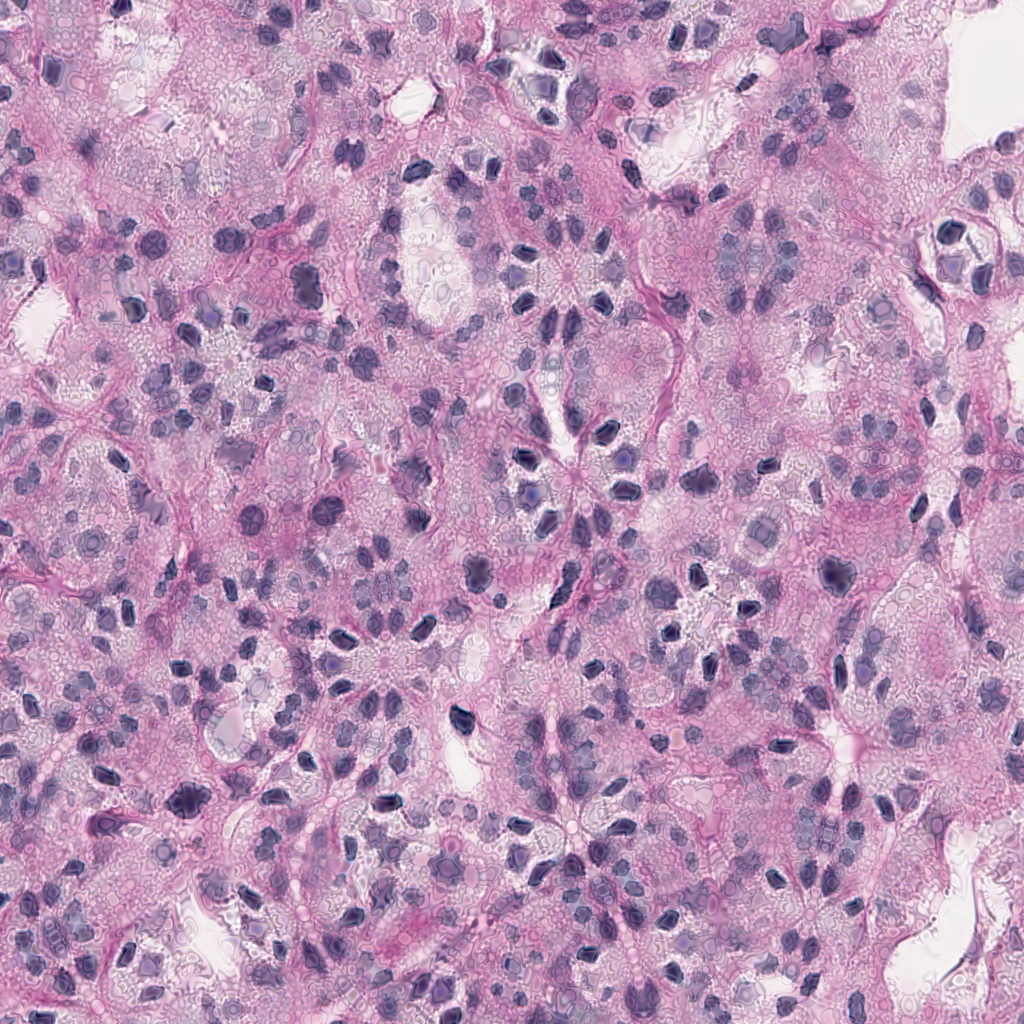}& 
    \includegraphics[height=\x]{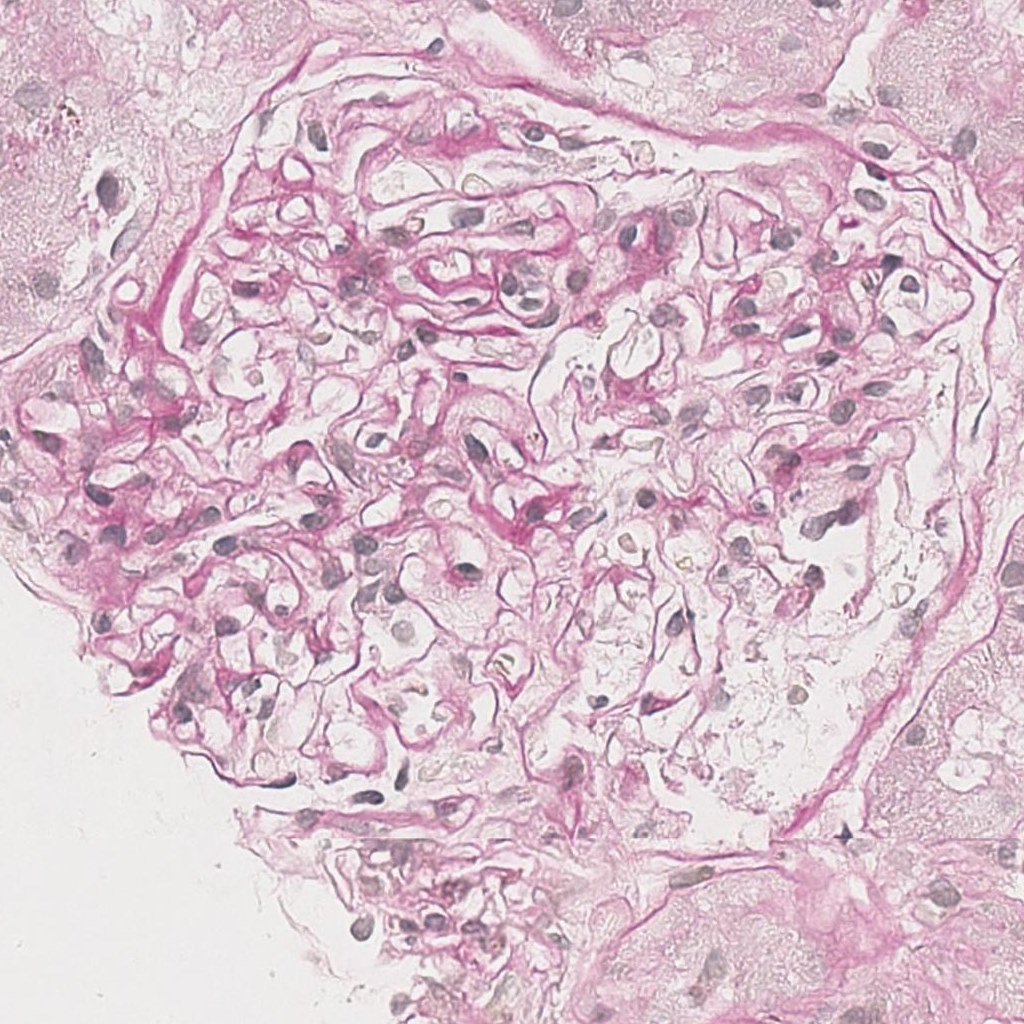} \\
    & \tiny{mag. 0} & \tiny{mag 0} & &  \tiny{mag. 2} & \tiny{mag. 2} \\
    & \tiny{StyleGAN} & \tiny{\ac{URCDM}}  && \tiny{Real} & \tiny{\ac{URCDM}}\\
     \raisebox{1.0cm}{\rotatebox[origin=l]{90}{\tiny{GLIOMA}}} &
    \includegraphics[height=\x]{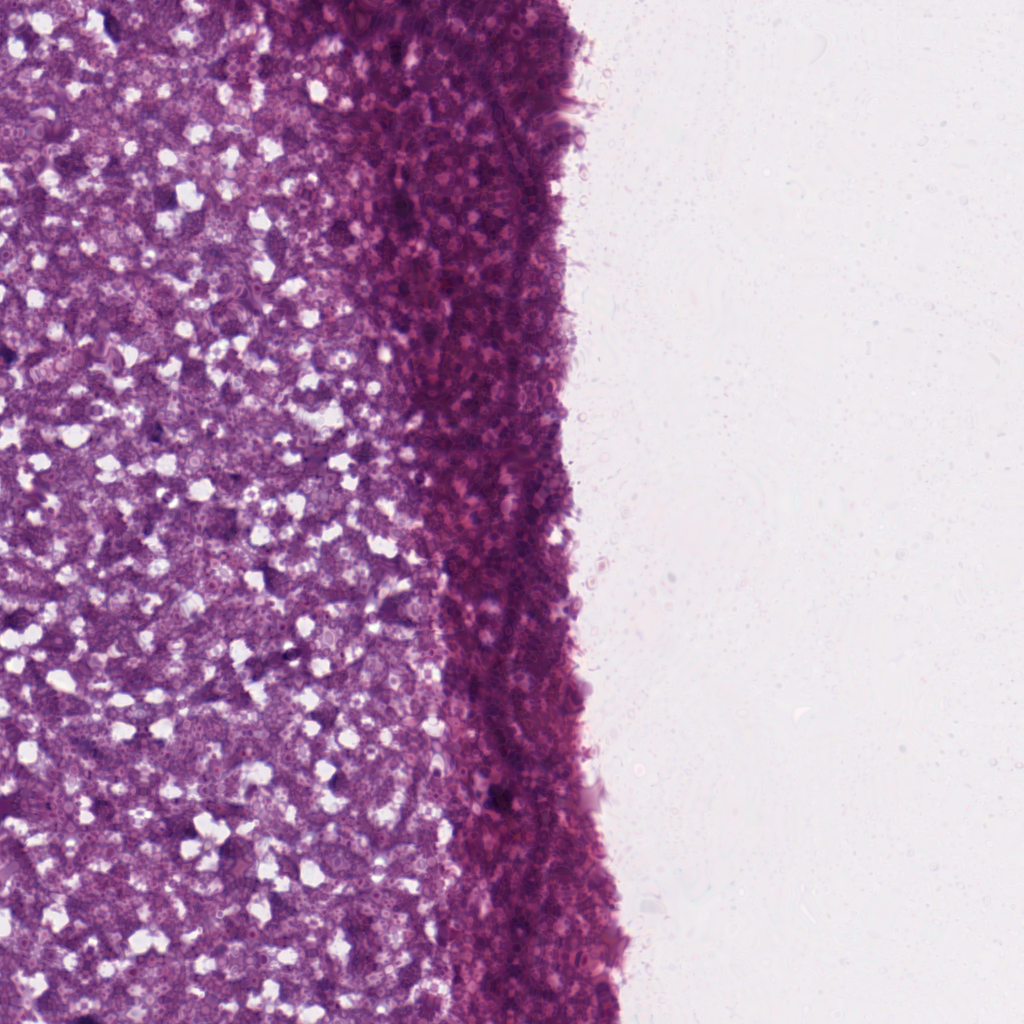}&
    \includegraphics[height=\x]{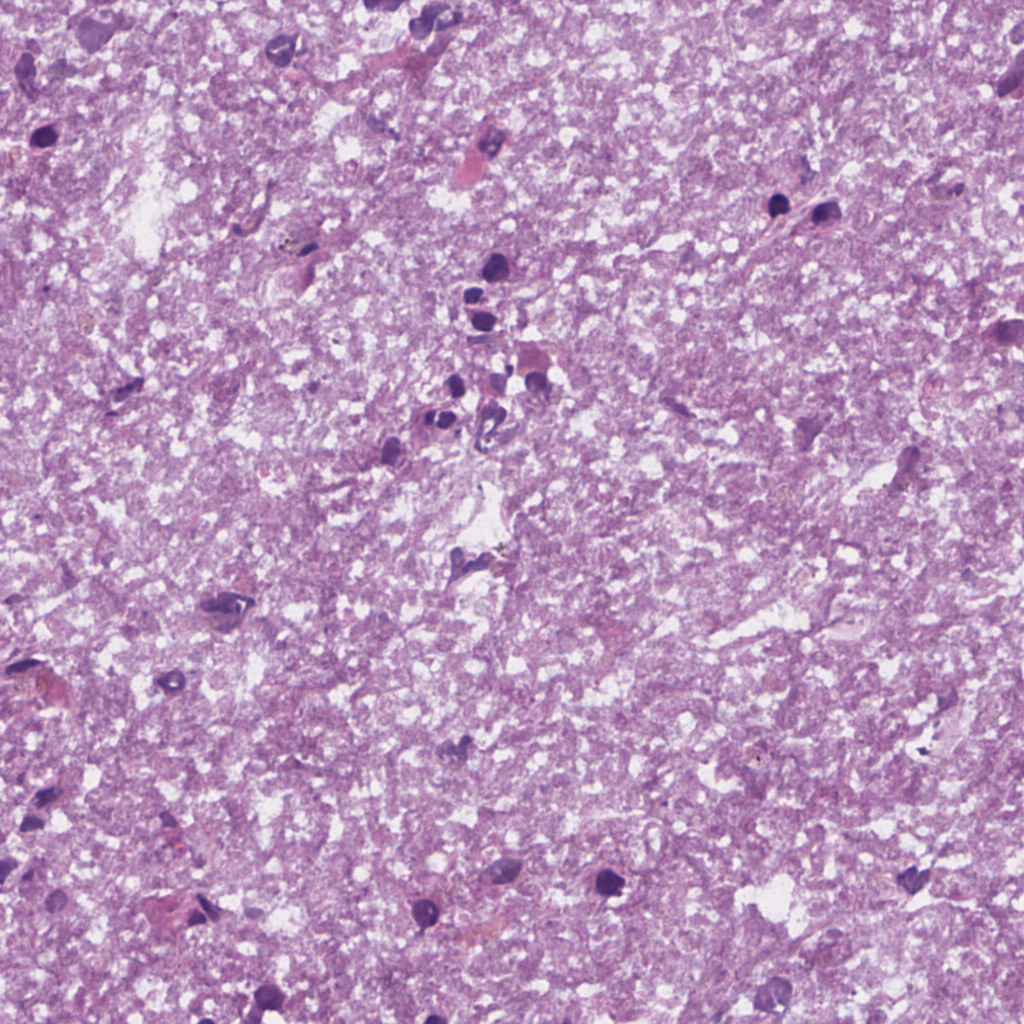}& 
    \raisebox{1.0cm}{\rotatebox[origin=l]{90}{\tiny{BREAST}}}  & 
    \includegraphics[height=\x]{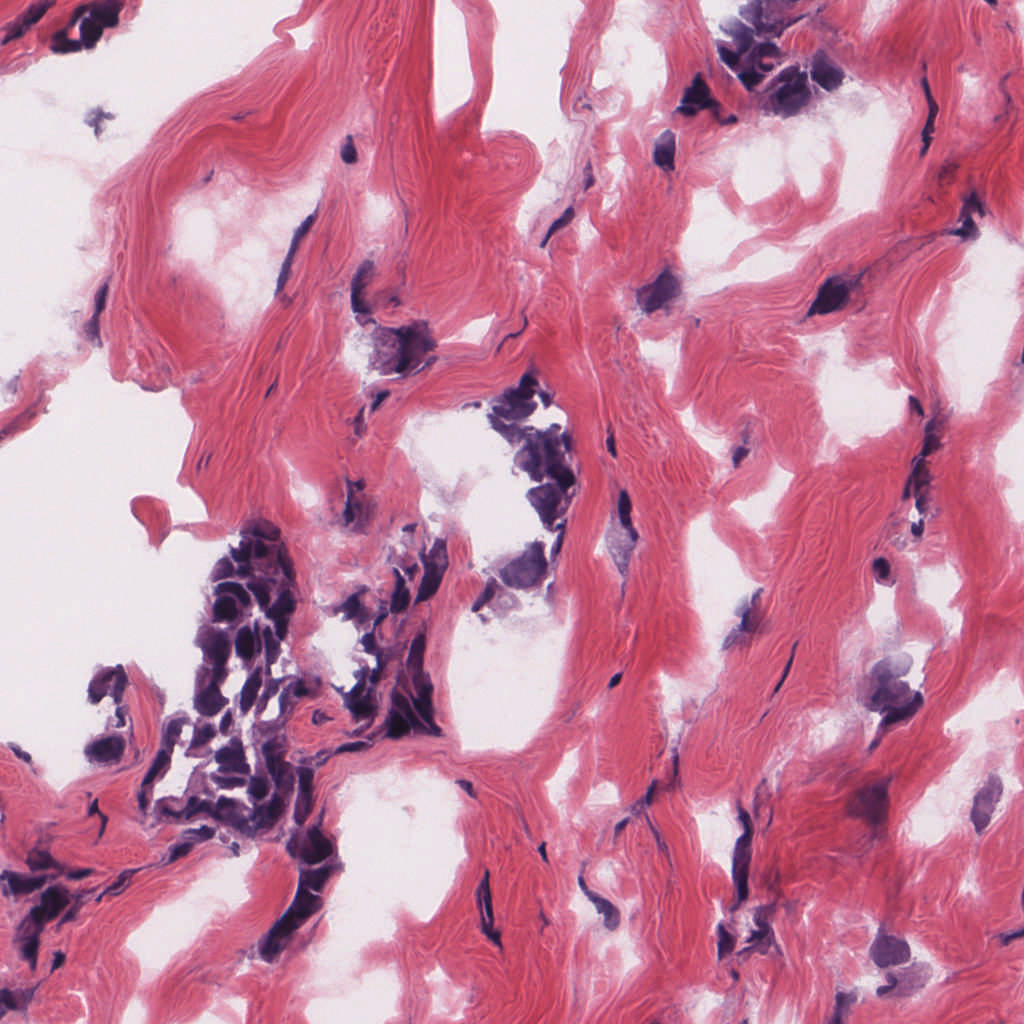}&
    \includegraphics[height=\x]{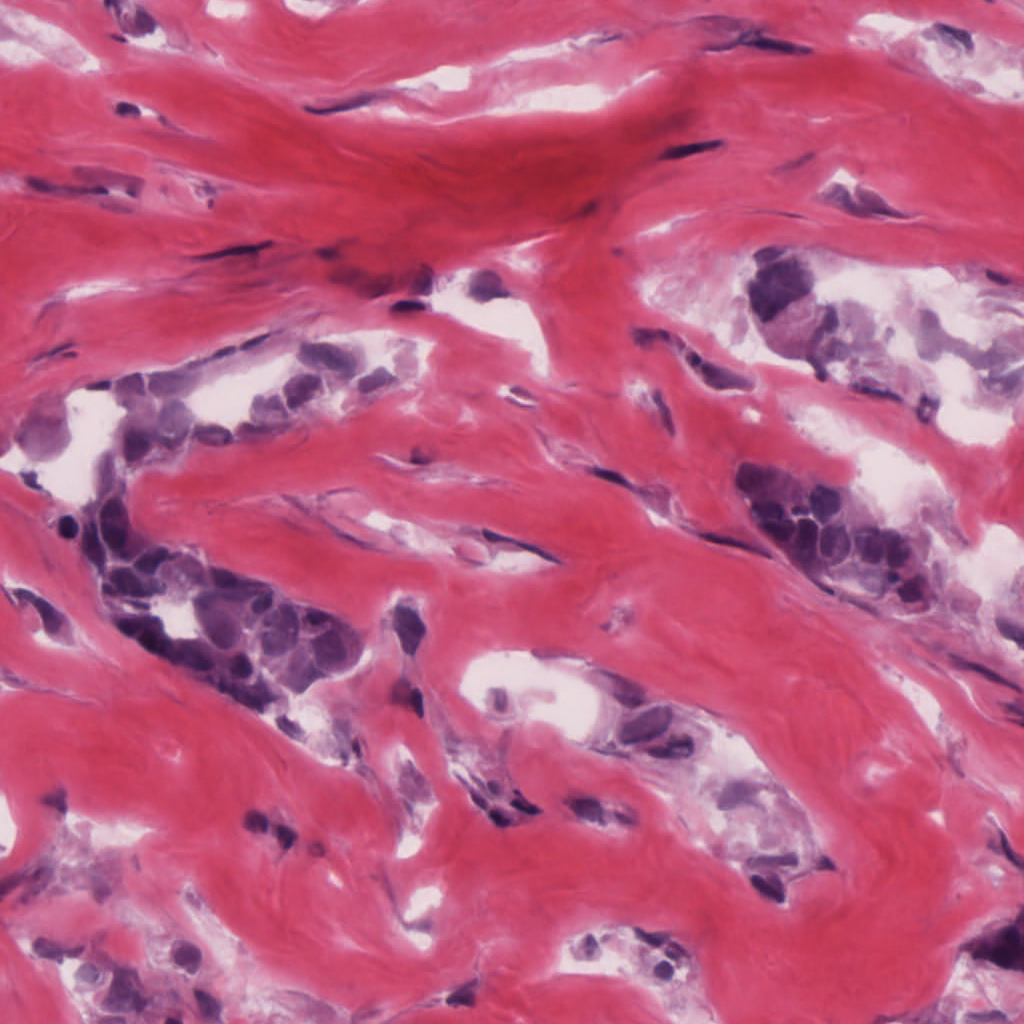}\\
     & \tiny{mag. 2} & \tiny{mag. 2} && \tiny{mag. 2} & \tiny{mag. 2} \\
\end{tabular}
\caption{Random \ac{URCDM} samples vs. Outpainting and StyleGAN. More qualitative examples can be found in the Appendix. \label{fig:urcdm-kidney}}
\end{figure}

\noindent\textbf{Human perception study:} 
To assess the perceived realism of synthetic images, a blinded comparison platform was developed, presenting expert pathologists with pairs of real and synthetic images from the KIDNEY dataset, and tasking them with identifying the authentic one. Results are shown in Table~\ref{tab:human-eval}. A $p$-value close to 0.5 means that the users are essentially guessing. Pathologists 2 and 4 were familiar with our approach and found a shortcut during high magnification evaluation. They looked for rare sample preparation artefacts, like tissue folding, which are not present in \ac{URCDM} since these parts are discarded during regular pathology assessment. 
In Table~\ref{tab:human-eval} the proportion of incorrectly classified samples $p$ varies considerably between users, with some having a strong preference for real images, and others having a strong preference for fake images. This balances out with the total proportion being close to $0.5$ for \ac{LRDM}. 

\noindent\textbf{Discussion}
\ac{URCDM} presents a shift from outpainting-based approaches. At a low resolution \ac{LRDM} outperform both StyleGans and MorphDiff~\cite{morphdiff}. This is likely due to \ac{GAN}s often causing unnatural symmetries to appear in the images, including ringing artefacts (see Figure \ref{fig:urcdm-kidney} KIDNEY/StyleGAN). The differential learning strategy employed by \ac{CDM}s, where distinct networks incrementally learn at each stage of the cascade, offers a tangible advantage over MorphDiff~\cite{morphdiff}.

Occasionally, at high magnification factors, isolated details of the images generated using outpainting are subjectively of higher quality, however, the lack of spatial coherency leads to poor pFID, IP and IR scores. 
The texture of the tissue for the KIDNEY dataset in Table~\ref{fig:urcdm-kidney} is realistically wispy in regions of the kidney further from the cortex, like in a real \ac{WSI}, the BREAST and GLIOMA datasets also follow similar spacial alignments. Additionally, structures that are only visible at high magnification are of high quality, like the glomerulus and tubules in the KINDEY or the smooth tissue surrounded by nuclei in BREAST in magnification~2 of Figure~\ref{fig:urcdm-kidney}.  
For trained pathologists, some fine details in the high-resolution image of Figure~\ref{fig:urcdm-kidney} can be poorer than those in the real \ac{WSI}s, which is likely due to \ac{URCDM}s having to maintain consistency through all nine stages. Base U-Nets are required to consistently and accurately `zoom in' the lower magnification image, and failure to do so will result in visually diverging neighbouring patches.  
Moreover, the performance analysis in Table~\ref{tab:ipir} illustrates a disparity in image quality across varying magnifications, underscoring a specific challenge in generating mid-magnification images. This disparity is primarily attributed to the constrained sampling process for magnification level 1, where the next stage relies on fully generated images, limiting the diversity and potentially the realism of the generated images. The unique challenge presented by the BREAST dataset, where larger image sizes mean that the $\num{40000} \times \num{40000}$ centre-crop may frequently capture non-informative white space, highlights the impact of dataset characteristics on the model's performance; this also lead to StyleGAN failing to learn the distribution.

\section{Conclusion}
\ac{URCDM}s are a novel way of generating images with more than $10^9$~pixels. Images generated by \ac{URCDM}s are spatially coherent over long distances and are plausible at different scales.
Fine details remain clear and are more coherent than images from outpainting, which is crucial for ultra-resolution imagery applications that use various image scales~\cite{yolt}.
Future work will focus on computational efficiency, the use of the same \ac{CDM} for all magnifications, and multi-modal learning.

\noindent\textbf{Acknowledgments:}
S. Cechnicka is supported by the UKRI Centre for Doctoral Training AI4Health  (EP / S023283/1). Support was also received from the ERC project MIA-NORMAL 101083647, the State of Bavaria (HTA) and DFG 512819079. HPC resources were provided by NHR@FAU of FAU Erlangen-N\"urnberg under the NHR project b180dc. NHR@FAU hardware is partially funded by the DFG – 440719683.
Dr. Roufosse is supported by the National Institute for Health Research (NIHR) Biomedical Research Centre based at Imperial College Healthcare NHS Trust and Imperial College London (ICL). The views expressed are those of the authors and not necessarily those of the NHS, the NIHR or the Department of Health. Dr Roufosse’s research activity is made possible with generous support from Sidharth and Indira Burman.
Human samples used in this research project were obtained from the Imperial College Healthcare Tissue \& Biobank (ICHTB). ICHTB is supported by NIHR Biomedical Research Centre based at Imperial College Healthcare NHS Trust and ICL. ICHTB is approved by Wales REC3 to release human material for research (22/WA/2836)

\noindent\textbf{\discintname: }
The authors have no competing interests to declare that are relevant to the content of this article.

\bibliographystyle{splncs04}

\newpage
\appendix

\section{Appendix}
\begin{figure}[htbp]
\centering
\caption{Samples of an ultra-resolution \ac{WSI}s generated using a \ac{URCDM}. Full-scale ultra-resolution image ($\num{41344} \times \num{41344}$ pixels), and highly zoomed-in crops of that image ($1024 \times 1024$ pixels) compared against random real and baseline images.}
\label{fig:urcdm-kidney_appendix}
\hspace{5em}\textbf{Magnification 0} \hspace{10em} \textbf{Magnification 2} \hfill
\newline
\rotatebox[origin=c]{90}{KIDNEY}\hspace{1mm} 
\begin{minipage}{.95\textwidth}
\begin{subfigure}{.15\textwidth}
  \caption*{Real}
  \includegraphics[width=\linewidth]{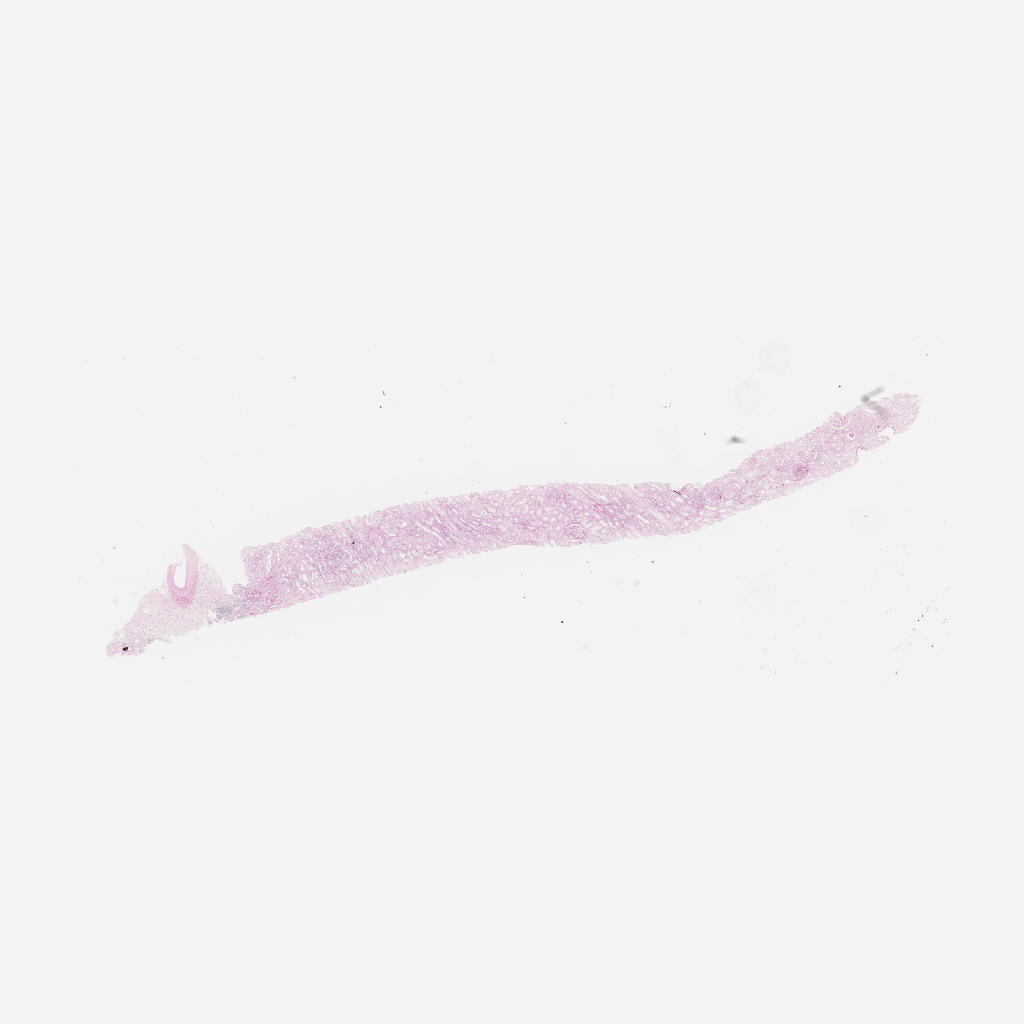}
\end{subfigure}\hfill
\begin{subfigure}{.15\textwidth}
  \caption*{URCDM}
  \includegraphics[width=\linewidth]{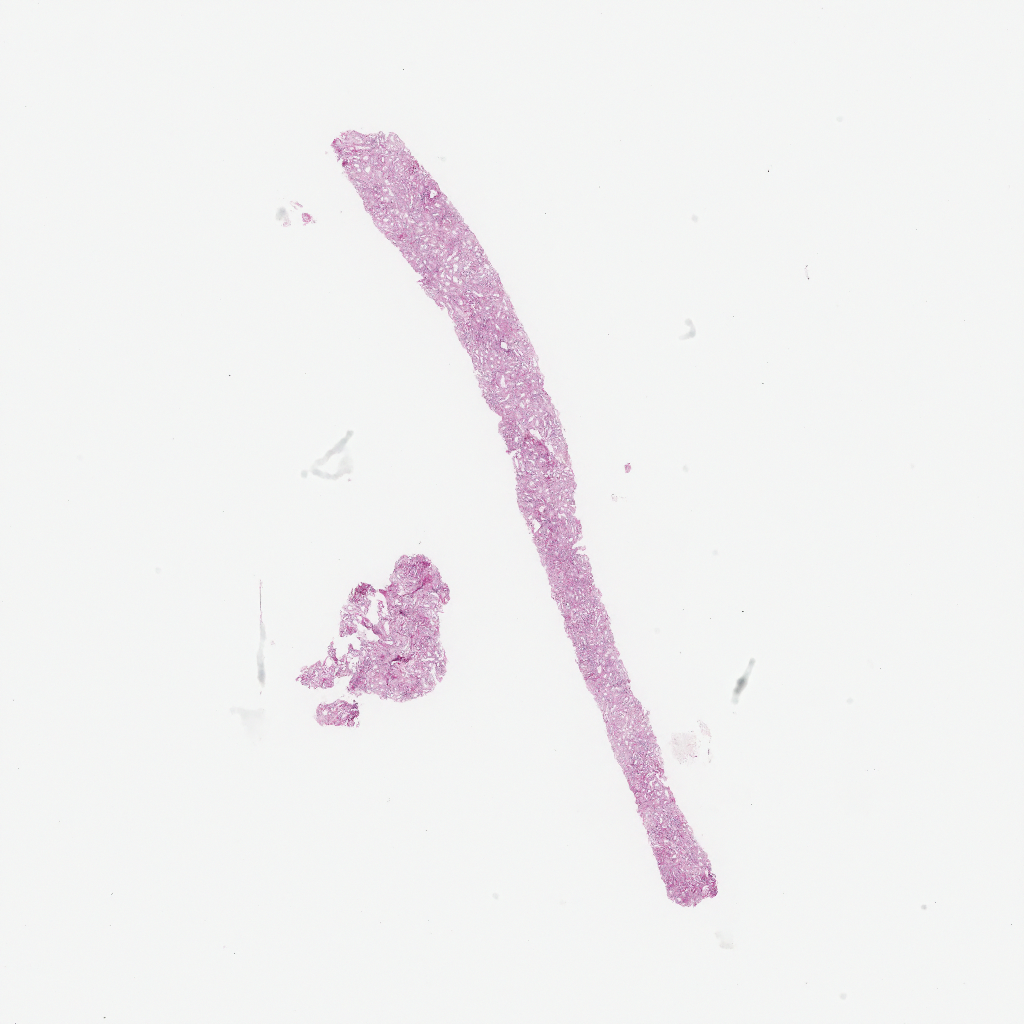}
\end{subfigure}\hfill
\begin{subfigure}{.15\textwidth}
  \centering
  \caption*{Outpainting}
  \includegraphics[width=\linewidth]{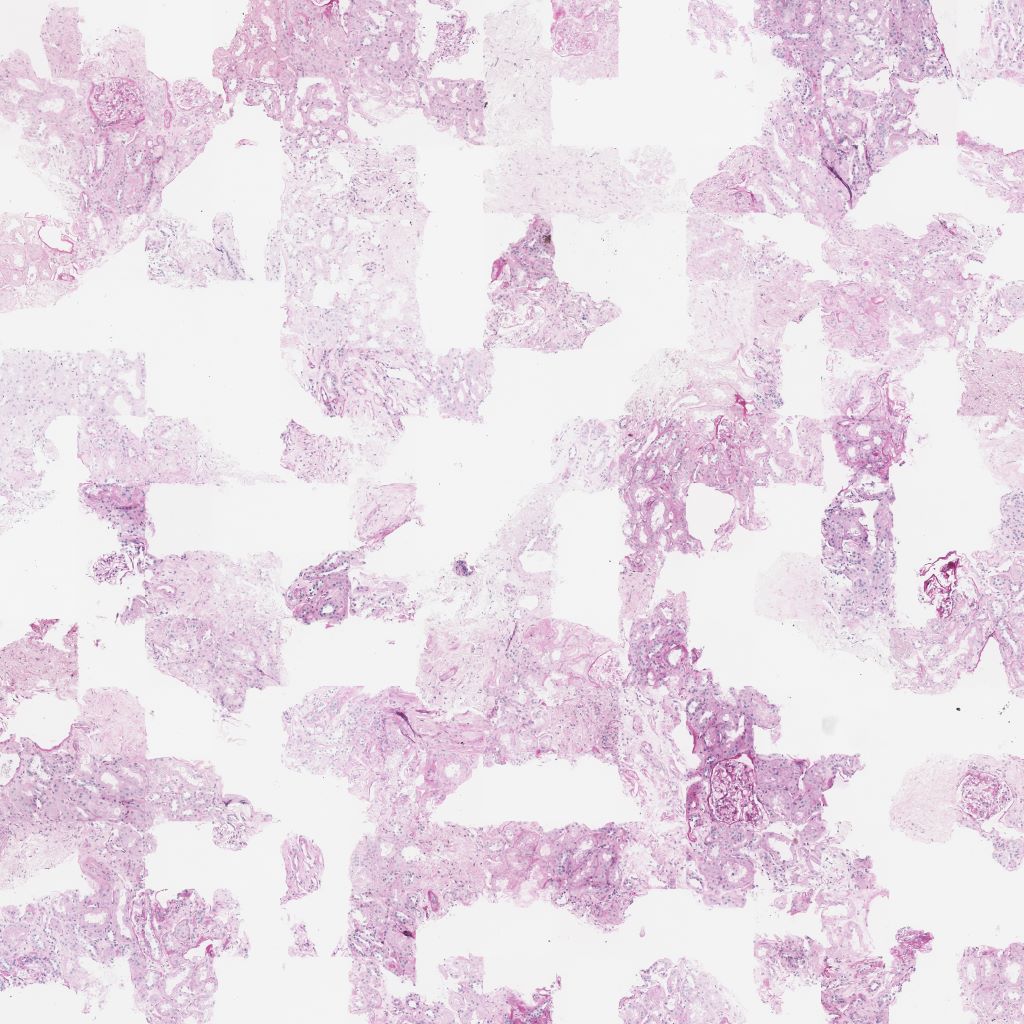}
\end{subfigure}
\begin{subfigure}{.02\textwidth}
  \centering
  \vrule width 1pt
\end{subfigure}\hfill
\begin{subfigure}{.15\textwidth}
  \caption*{Real}
  \includegraphics[width=\linewidth]{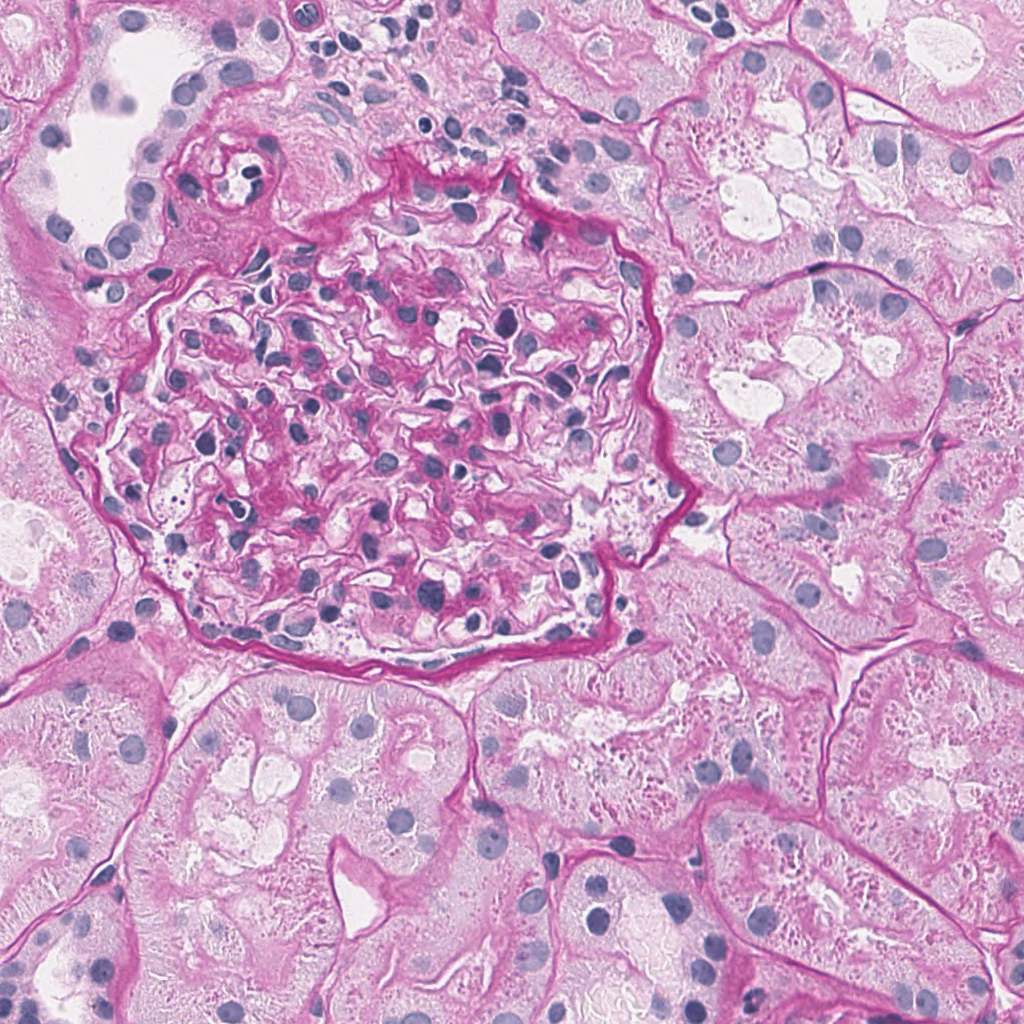}
\end{subfigure}\hfill
\begin{subfigure}{.15\textwidth}
  \caption*{URCDM}
  \includegraphics[width=\linewidth]{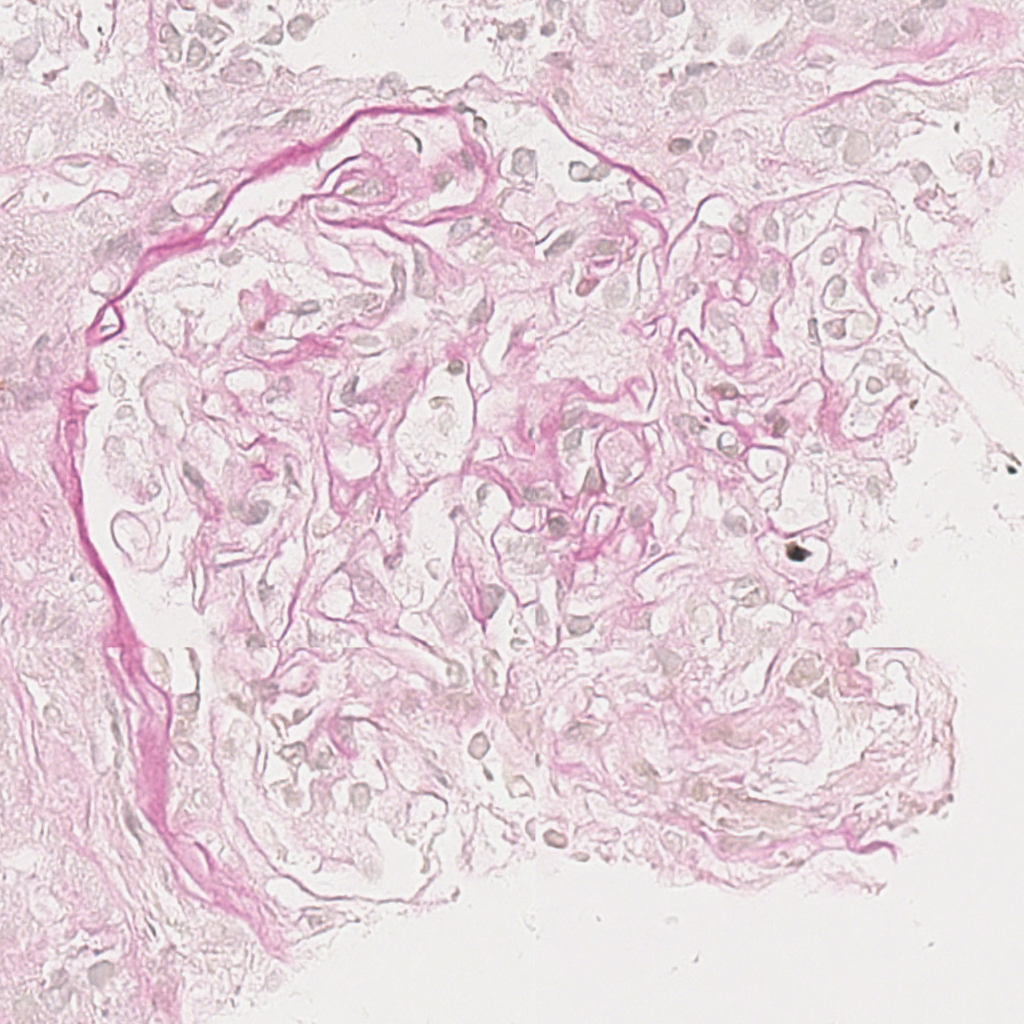}
\end{subfigure}\hfill
\begin{subfigure}{.15\textwidth}
  \caption*{StyleGAN}
  \includegraphics[width=\linewidth]{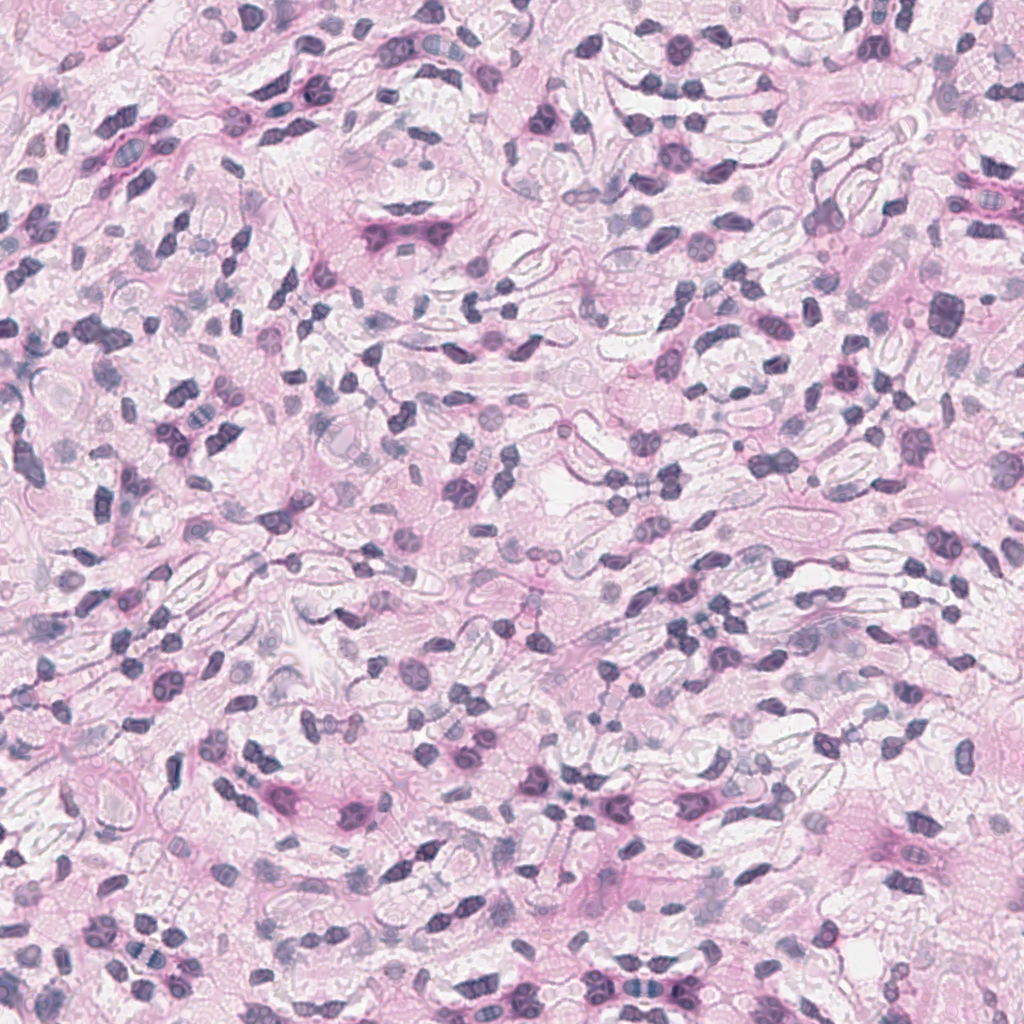}
\end{subfigure}\hfill
\end{minipage}
\vspace{1em}
\rotatebox[origin=c]{90}{GLIOMA}\hspace{1mm} 
\begin{minipage}{.95\textwidth}
    \begin{subfigure}{.15\textwidth}
      \includegraphics[width=\linewidth]{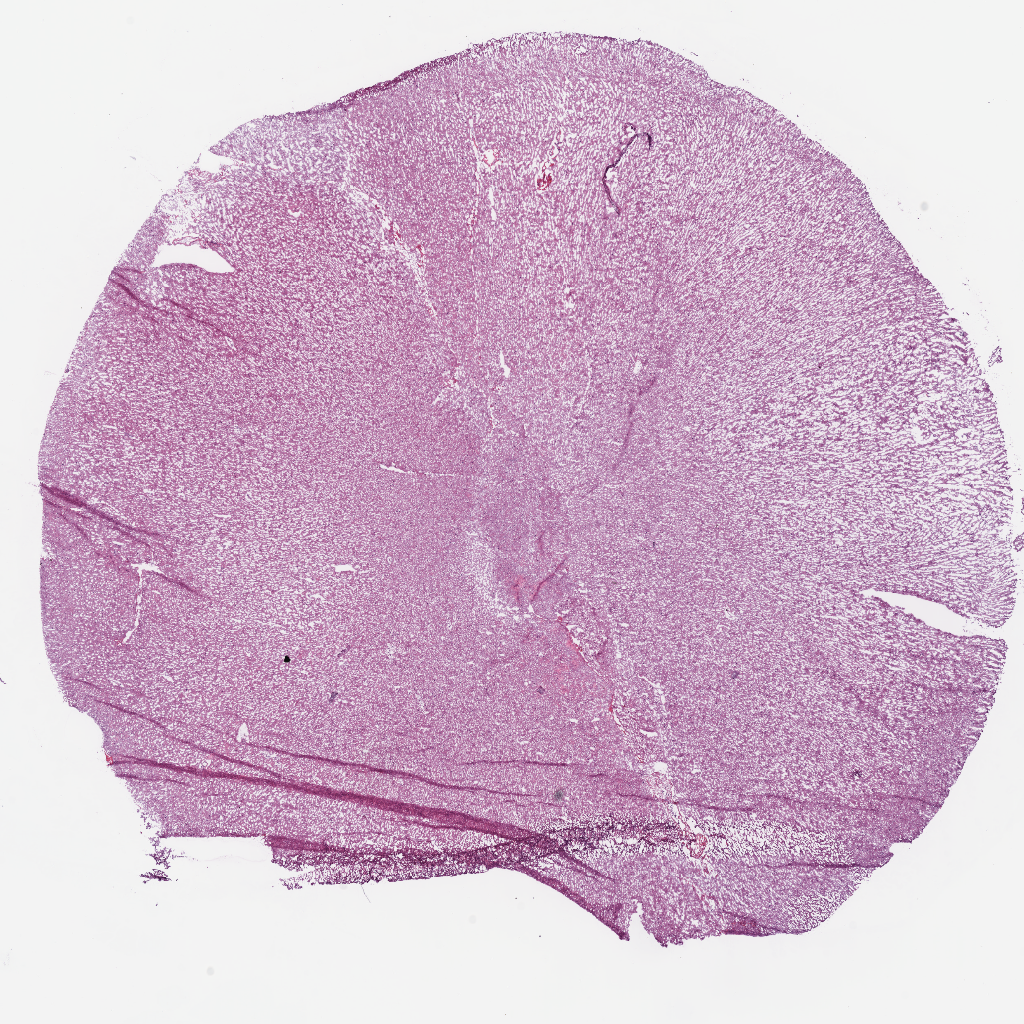}
    \end{subfigure}\hfill
    \begin{subfigure}{.15\textwidth}
      \includegraphics[width=\linewidth]{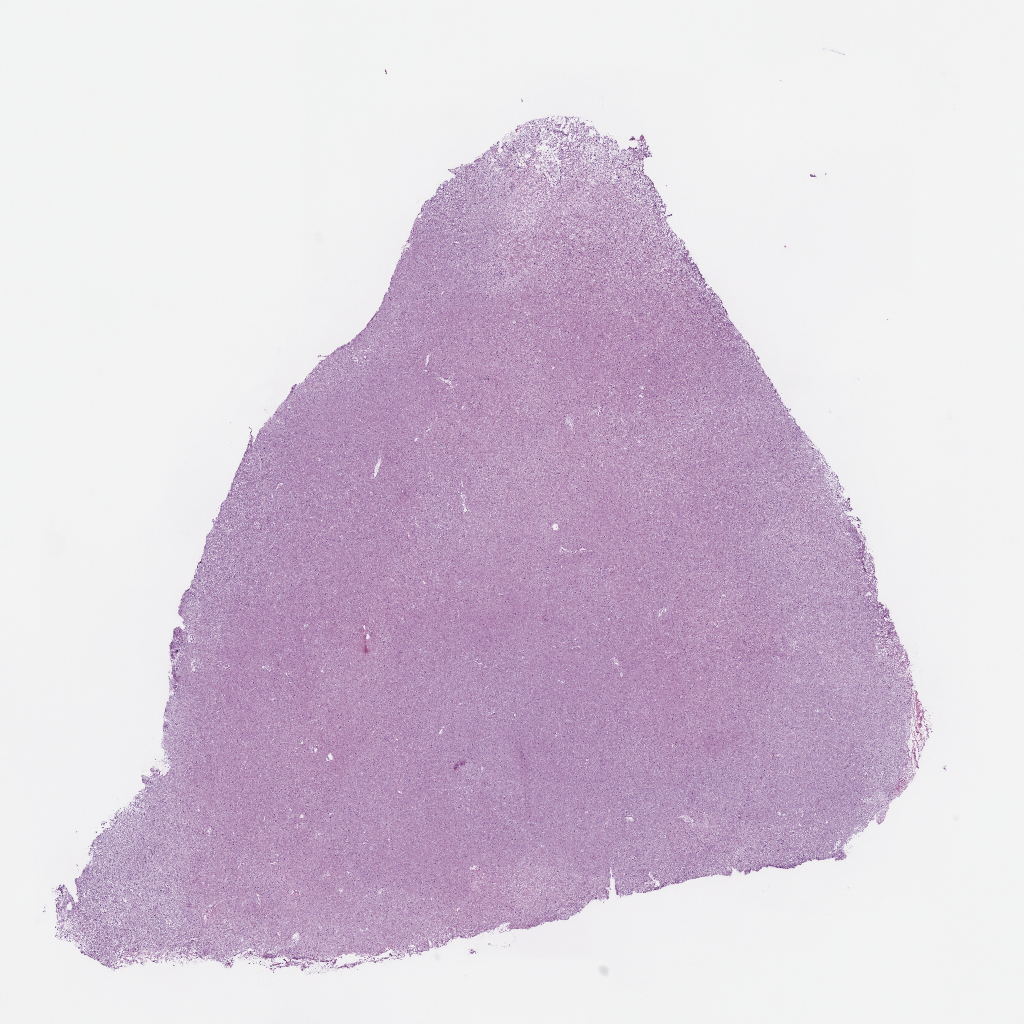}
    \end{subfigure}\hfill
    \begin{subfigure}{.15\textwidth}
      \centering
      \includegraphics[width=\linewidth]{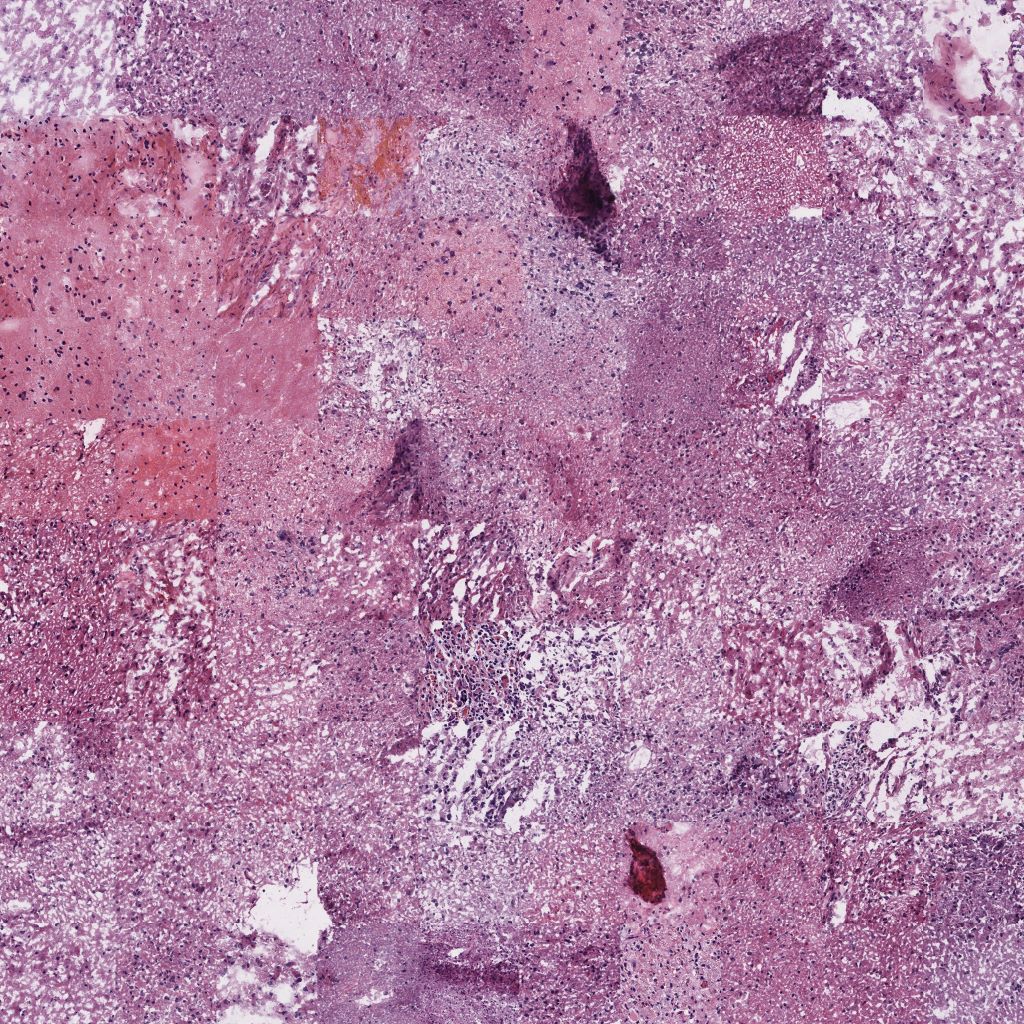}
    \end{subfigure}
    \begin{subfigure}{.02\textwidth}
      \centering
      \vrule width 1pt
    \end{subfigure}\hfill
    \begin{subfigure}{.15\textwidth}
      \includegraphics[width=\linewidth]{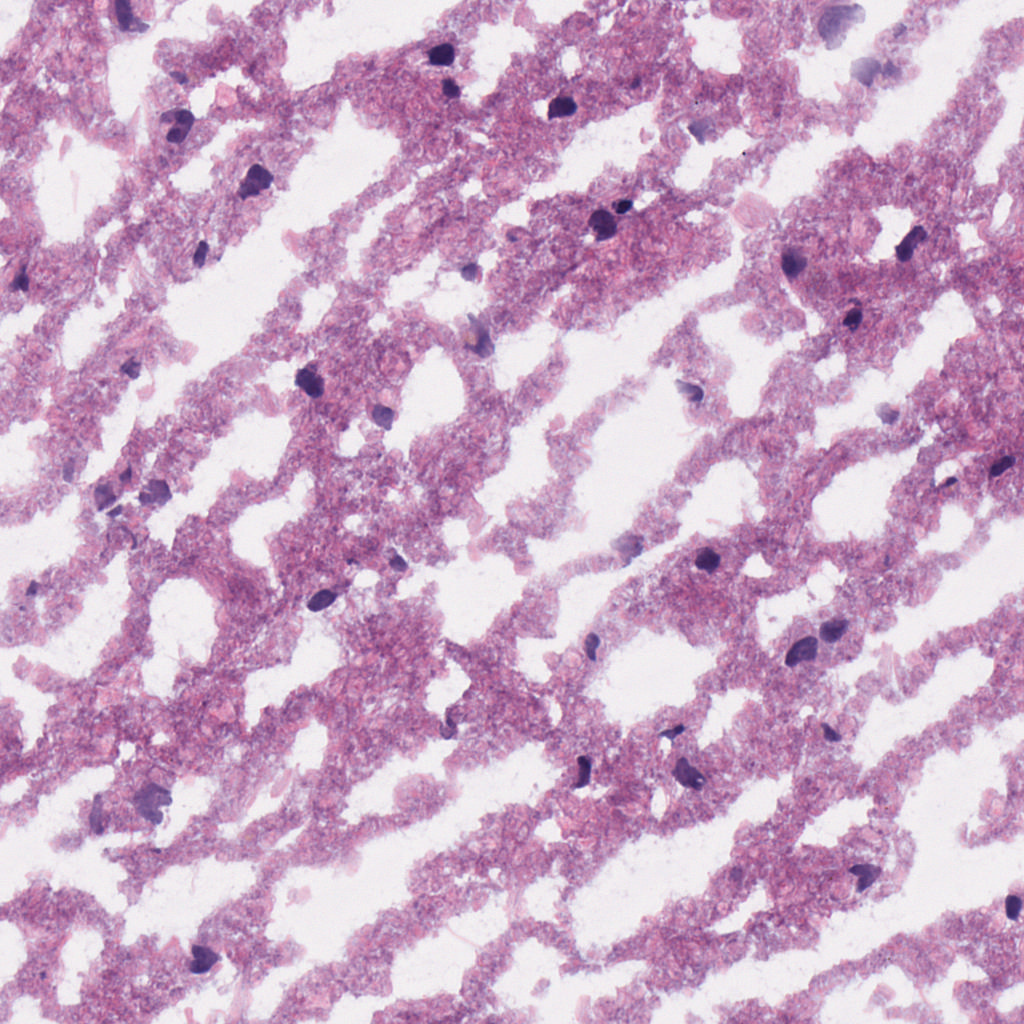}
    \end{subfigure}\hfill
    \begin{subfigure}{.15\textwidth}
      \includegraphics[width=\linewidth]{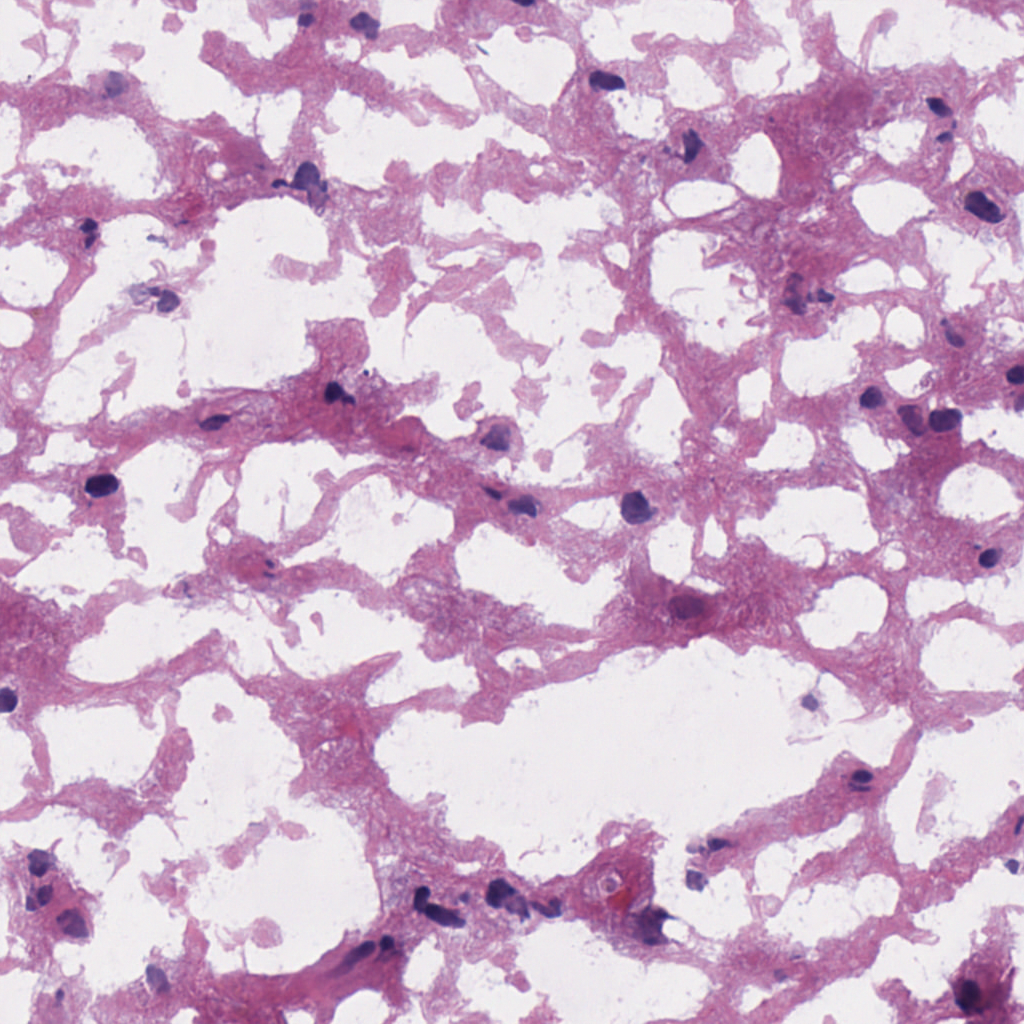}
    \end{subfigure}\hfill
    \begin{subfigure}{.15\textwidth}
      \includegraphics[width=\linewidth]{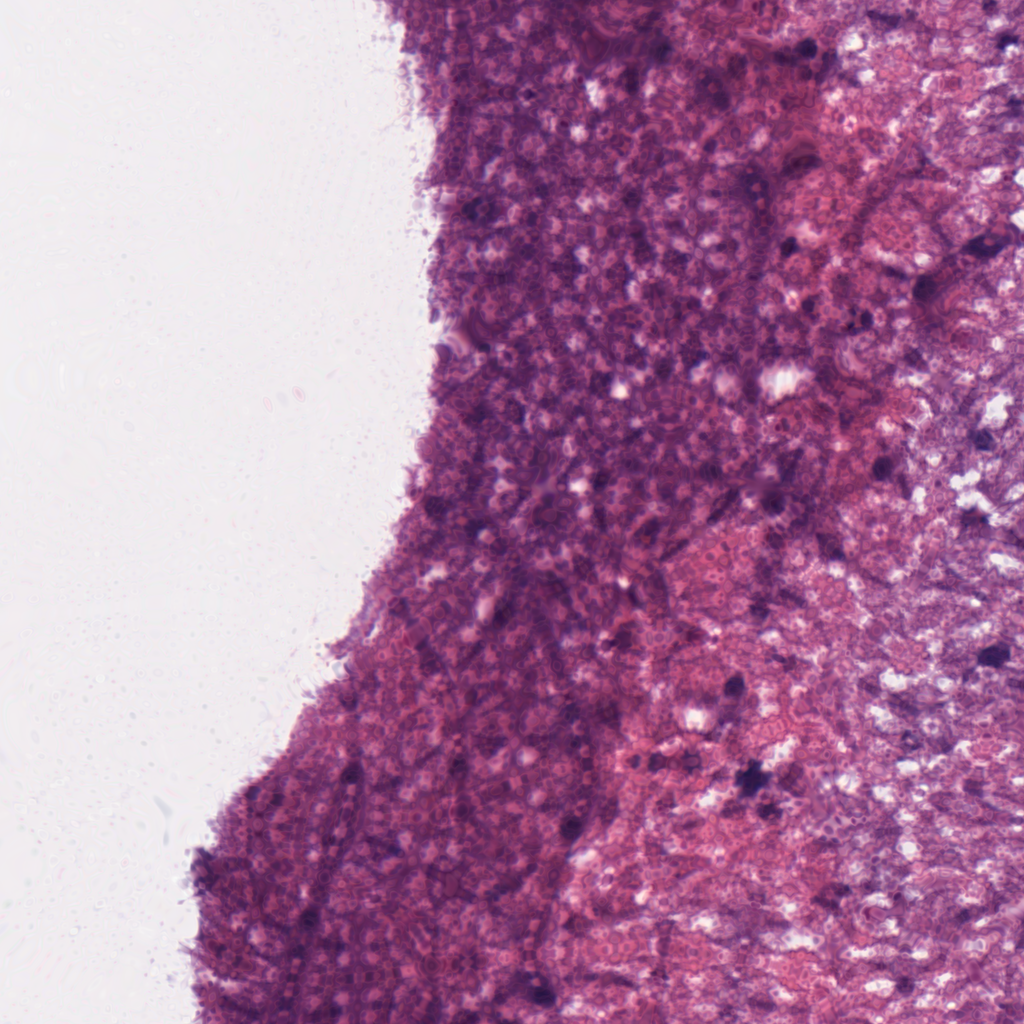}
    \end{subfigure}\hfill
\end{minipage}

\rotatebox[origin=c]{90}{BREAST}\hspace{1mm} 
\begin{minipage}{.95\textwidth}
\begin{subfigure}{.15\textwidth}
  \includegraphics[width=\linewidth]{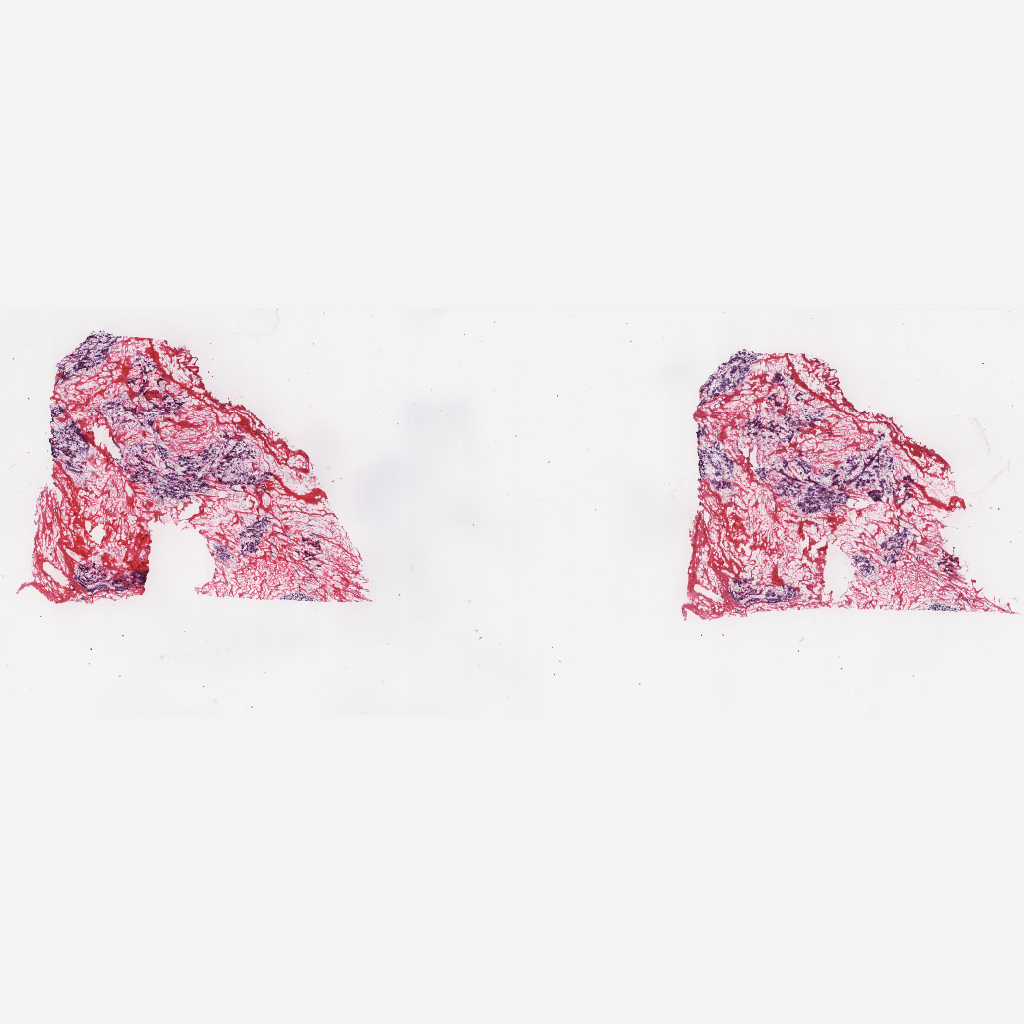}
\end{subfigure}\hfill
\begin{subfigure}{.15\textwidth}
  \includegraphics[angle=90,width=\linewidth]{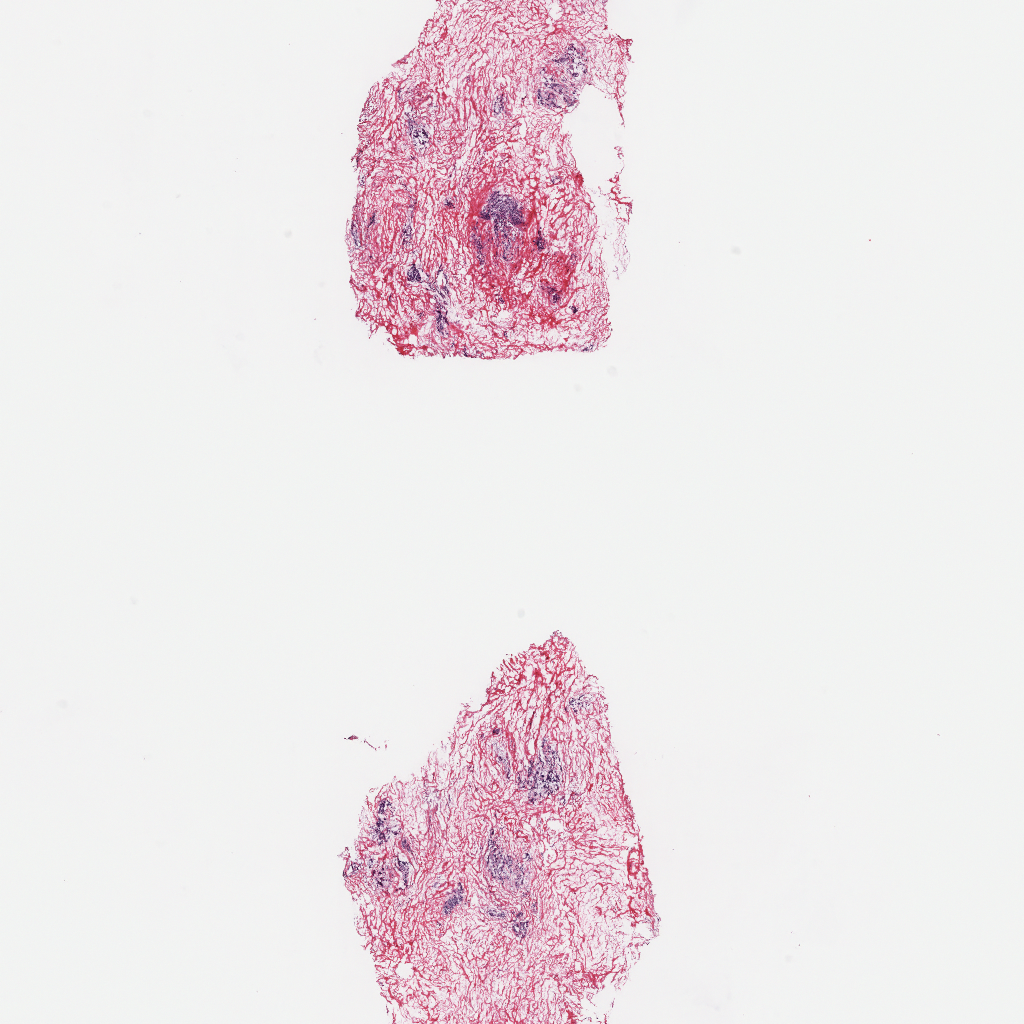}
\end{subfigure}\hfill
\begin{subfigure}{.15\textwidth}
  \centering
  \includegraphics[width=\linewidth]{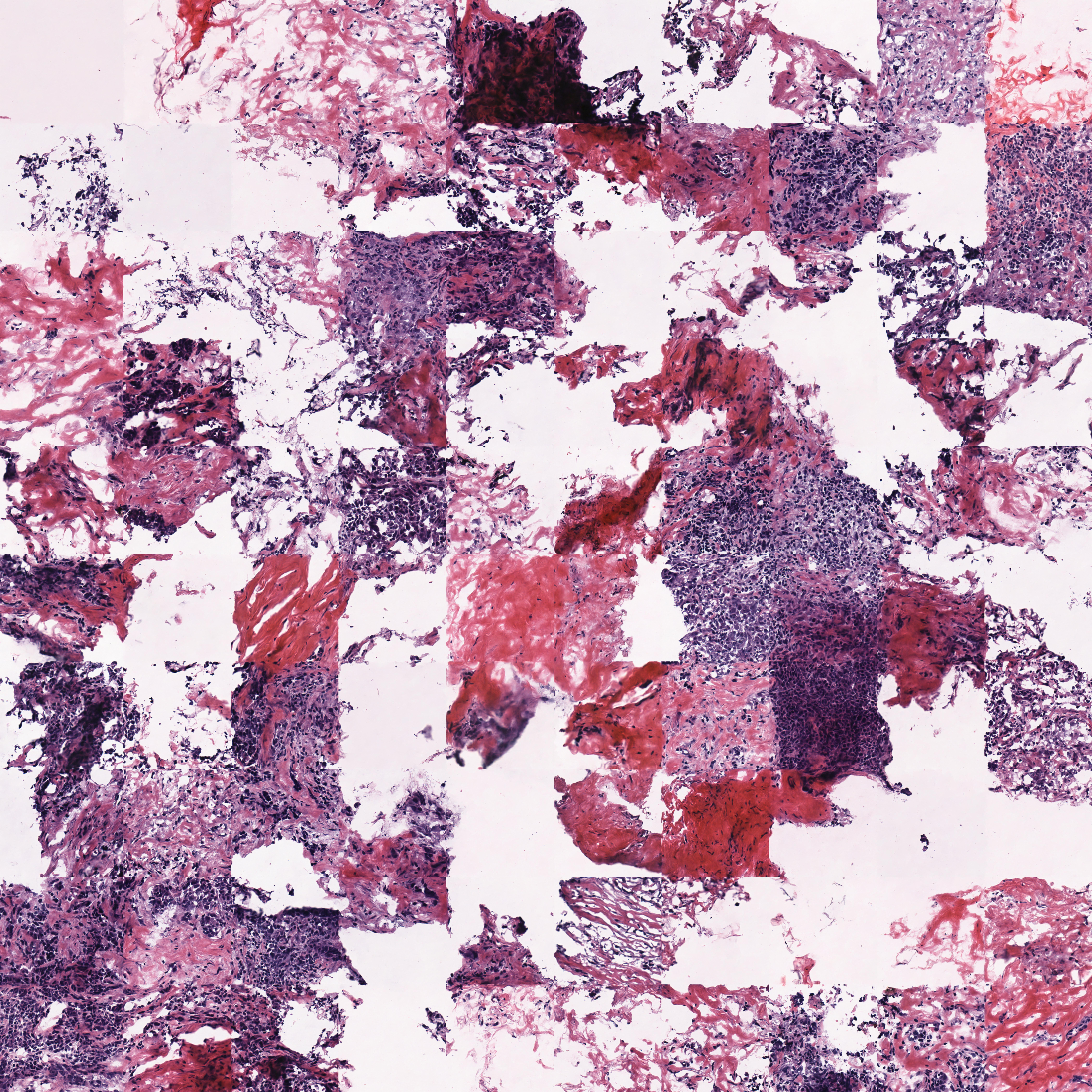}
\end{subfigure}
\begin{subfigure}{.02\textwidth}
  \centering
  \vrule width 1pt
  \caption*{}
\end{subfigure}\hfill
\begin{subfigure}{.15\textwidth}
  \includegraphics[angle=90,width=\linewidth]{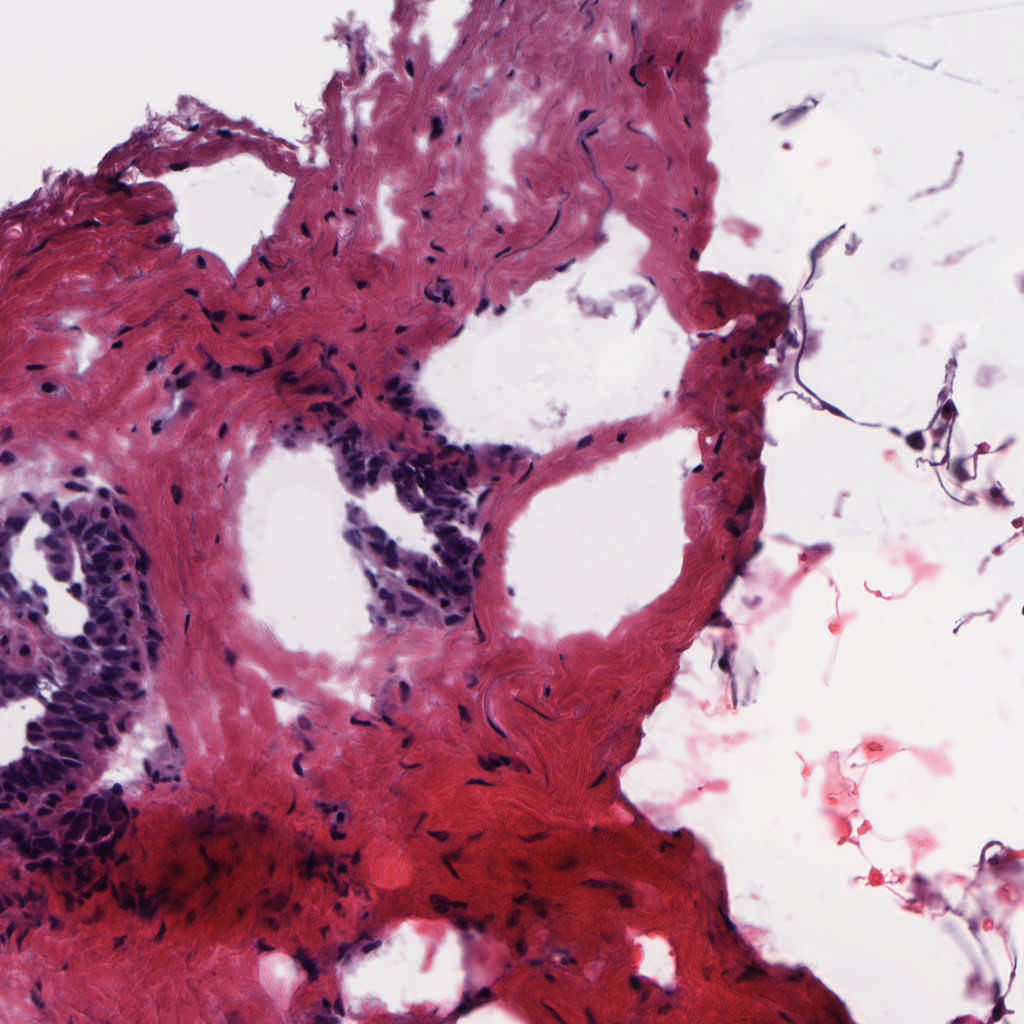}
\end{subfigure}\hfill
\begin{subfigure}{.15\textwidth}
  \includegraphics[width=\linewidth]{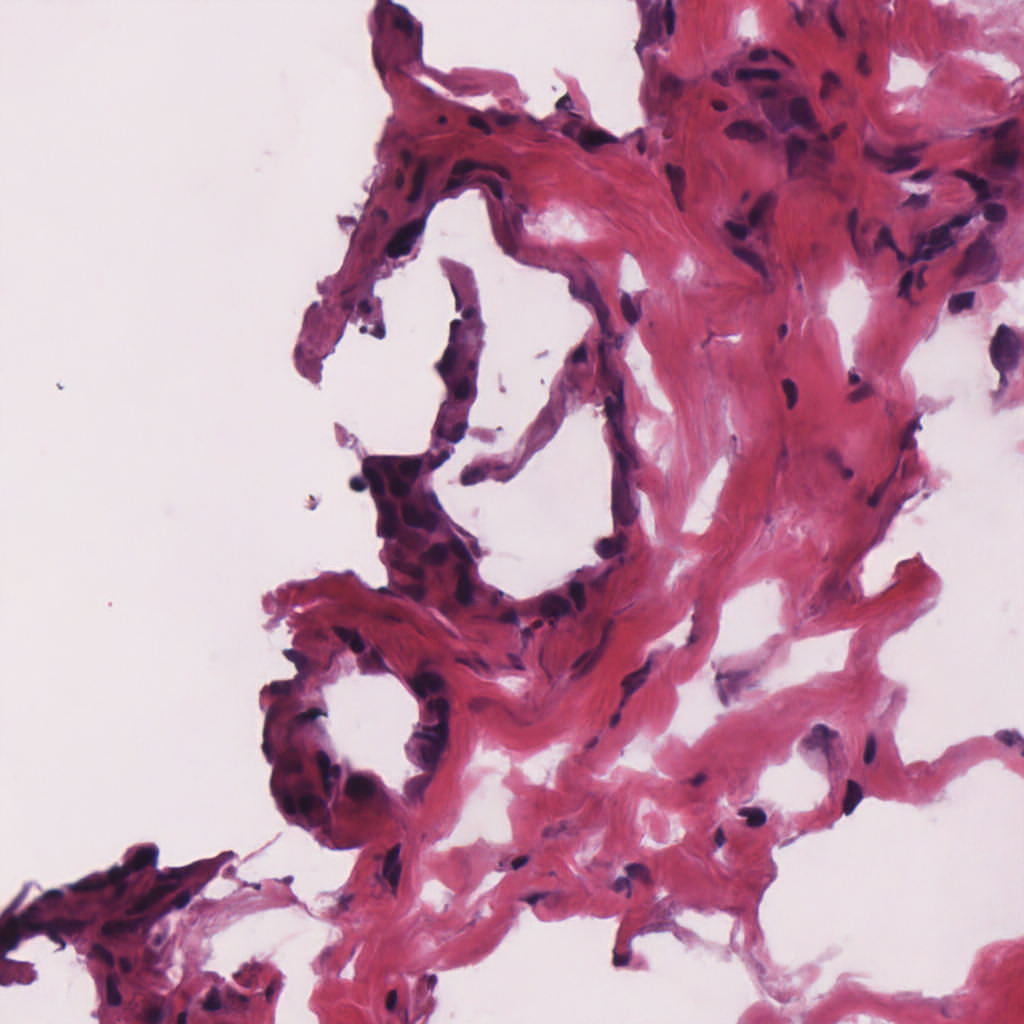}
\end{subfigure}\hfill
\begin{subfigure}{.15\textwidth}
  \includegraphics[width=\linewidth]{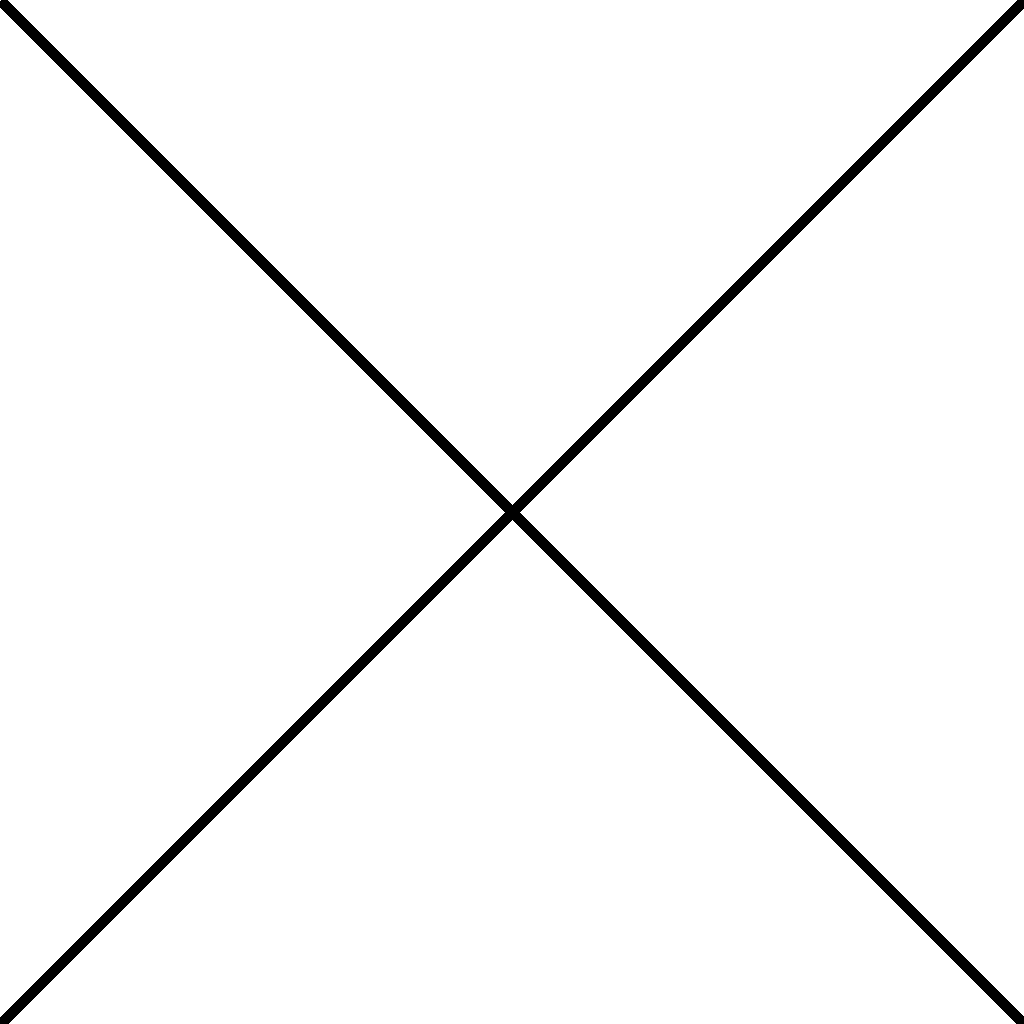}
\end{subfigure}
\end{minipage}
\end{figure}
\end{document}